\documentclass{article}
\usepackage{amsmath}
\usepackage{amssymb}
\usepackage{epsfig}
\usepackage{color}

\makeatletter
\@addtoreset{equation}{section}

\makeatother

\begin{document}

\begin{flushright}
June 2011

SNUTP11-004
\end{flushright}

\begin{center}

\vspace{5cm}

{\LARGE 
\begin{center}
Eigenvalue Distributions in Matrix Models for Chern-Simons-matter Theories
\end{center}
}

\vspace{2cm}

Takao Suyama \footnote{e-mail address : suyama@phya.snu.ac.kr}

\vspace{1cm}

{\it 
BK-21 Frontier Research Physics Division

and 

Center for Theoretical Physics, 

\vspace{2mm}

Seoul National University, 

Seoul 151-747 Korea}

\vspace{3cm}

{\bf Abstract} 

\end{center}

The eigenvalue distribution is investigated for matrix models related via the localization to Chern-Simons-matter 
theories. 
An integral representation of the planar resolvent is used to derive the positions of the branch points of the planar resolvent 
in the large 't~Hooft coupling limit. 
Various known exact results on eigenvalue distributions and the expectation value of Wilson loops are reproduced. 

\newpage

\vspace{1cm}

\section{Introduction}

\vspace{5mm}

Superconformal Chern-Simons-matter theories have attracted much attention recently in relation to the dynamics of M2-branes. 
The worldvolume theory on M2-branes was proposed in 
\cite{Bagger:2006sk}\cite{Bagger:2007jr}\cite{Bagger:2007vi}\cite{Gustavsson:2008dy}\cite{Gustavsson:2007vu} 
based on a 3-algebra structure. 
The theory turned out to be equivalent to a Chern-Simons theory with a product gauge group coupled to matters 
\cite{VanRaamsdonk:2008ft}. 
Chern-Simons-matter theories with product gauge groups were also studied in \cite{Gaiotto:2008sd} in a different context, and 
then the construction of \cite{Gaiotto:2008sd} was extended to more general theories \cite{Hosomichi:2008jd} (see also 
\cite{Hosomichi:2008jb}). 
Soon after those developments, 
the worldvolume theory on any number $N$ of M2-branes placed at the tip of an orbifold was constructed in \cite{Aharony:2008ug}. 
Since this theory, known as ABJM theory, allows us to take the large $N$ limit, one may discuss its gravity dual, as in the case 
of ${\cal N}=4$ super Yang-Mills theory in four dimensions \cite{Maldacena:1997re}. 
However, as usual in AdS/CFT correspondence, it is very difficult to check the correspondence between the field theory and 
the dual gravity or M-theory, unless it is possible to perform an exact calculation in the field theory 
side, such as a calculation of a quantity which is protected by the BPS nature or the integrability. 

Fortunately, there exists a technique, known as the localization, which is quite powerful for calculating 
some class of observables in field theories. 
The localization technique may be used to calculate expectation values of operators which preserve a particular supercharge. 
The reason that this technique is so powerful is that it enables one to perform a path-integral exactly by just evaluating 
the saddle-points. 
In the context of AdS$_5$/CFT$_4$ correspondence, the localization technique was used in \cite{Pestun:2007rz} 
to calculate the expectation value of the half-BPS Wilson loop. 
As a result, one obtains a function of the 't~Hooft coupling $\lambda$ which reproduces, via the relation 
\cite{Rey:1998ik}\cite{Maldacena:1998im}\cite{Rey:1998bq}, 
the behavior of the string worldsheet in AdS$_5$ exactly in the large $\lambda$ limit. 

A similar technique was applied to Chern-Simons-matter theories in \cite{Kapustin:2009kz}. 
The localization formula obtained in \cite{Kapustin:2009kz} is much simpler than that for a generic four-dimensional 
${\cal N}=2$ gauge theory in 
\cite{Pestun:2007rz}. 
Therefore, the analysis using this formula can be done rather easily. 
First, the ordinary 't~Hooft limit of ABJM theory was investigated 
\cite{Suyama:2009pd}\cite{Marino:2009jd} using the localization formula, and the behavior of 
the BPS Wilson loop \cite{Marino:2009jd} constructed in 
\cite{Drukker:2008zx}\cite{Chen:2008bp}\cite{Rey:2008bh} and the free energy \cite{Drukker:2010nc}, expected from AdS/CFT 
correspondence, were reproduced exactly. 
Note that the dual theory for the 't~Hooft limit is Type IIA theory \cite{Aharony:2008ug}. 
Next, another limit, called the M-theory limit, was investigated in \cite{Herzog:2010hf}. 
This is a large $N$ limit with $k$ kept finite, so that the dual theory should be M-theory, not Type IIA theory. 
The technique developed in \cite{Herzog:2010hf} enabled the authors to derive the exact behavior of the free energy and the 
Wilson loops 
in the M-theory limit. 
Remarkably, this technique can be applied to various other Chern-Simons-matter theories whose planar resolvents were not known. 

The investigation based on the localization formula had been 
limited to the theories which have at least ${\cal N}=3$ supersymmetry, 
since otherwise the dimensions of the fundamental fields may be renormalized in the IR, 
and localization formula in \cite{Kapustin:2009kz} may not valid. 
Recently, the localization formula was generalized to ${\cal N}=2$ theories in \cite{Jafferis:2010un}\cite{Hama:2010av} 
which plays a central role in recent developments on Chern-Simons-matter theories, especially on the so-called ``F-theorem'' 
\cite{Jafferis:2010un}. 
The analyses in \cite{Herzog:2010hf} has been extended to ${\cal N}=2$ theories in 
\cite{Martelli:2011qj}\cite{Cheon:2011vi}\cite{Jafferis:2011zi}\cite{Niarchos:2011sn}\cite{Gulotta:2011si}. 
See also \cite{Santamaria:2010dm} for another approximation scheme in a similar spirit.  

\vspace{5mm}

In the 't~Hooft limit, one may deal with the planar resolvent, as in \cite{Marino:2009jd}. 
The planar resolvent is a standard machinery in the context of the matrix model. 
It contains all the information on the planar limit of the matrix model. 
The localization formula in \cite{Kapustin:2009kz} is given in terms of a finite-dimensional integral. 
Since it looks like the partition function of a matrix model, the techniques developed in the matrix model context may be 
applicable. 
In \cite{Marino:2009jd}, the planar resolvent was obtained for arbitrary value of the 't~Hooft coupling, and therefore, 
it is possible to obtain, for example, 
an interpolating function for the Wilson loop between the weak coupling region and the strong coupling region. 
Systematic expansions around the weak coupling limit and the strong coupling limit are also available. 
However, it is in general difficult to obtain the explicit 
formula of the planar resolvent for generic Chern-Simons-matter theories. 
In the case of ABJM theory, a relation to a topological string theory was efficiently used, but it seems to be an 
accidental relation since the planar resolvent of ABJM theory coupled to fundamental matters does not seem to allow such a 
relation \cite{Santamaria:2010dm}. 

On the other hand, the technique in \cite{Herzog:2010hf} is rather simple, and it can be applied to various Chern-Simons-matter 
theories. 
One may derive the large $N$ results directly from the matrix model action, without relying on the use of the planar resolvent. 
Up to some ansatz on the eigenvalue distributions, which is supported by numerical calculations, the technique provides us with 
the configuration of the eigenvalues as well as the density distribution of the eigenvalues. 
Therefore, following \cite{Herzog:2010hf}, one may obtain all the information on the large $N$ limit of the theory. 
However, this technique allows one to obtain only the results in the large $N$ limit, and the systematic $1/N$ expansion is not yet 
available. 

\vspace{5mm}

In this paper, we present the third method to investigate matrix models related to Chern-Simons-matter theories. 
It is a first-principle calculation based on the planar resolvent, but it does not require the expression for the resolvent as 
explicit as in \cite{Marino:2009jd}. 
It was shown in \cite{Suyama:2010hr} that the planar resolvent of some Chern-Simons-matter theories can be written as contour 
integrals explicitly up to a few parameters. 
The equations which determine the parameters were also derived \cite{Suyama:2010hr}. 
We show that the integral representation of the planar resolvent is already useful enough 
to derive the results in the large 't~Hooft coupling limit. 
Moreover, it turns out that the derivation of those results is rather elementary; no use of the machinery of topological string 
theory is necessary. 
The systematic expansion in terms of an inverse power of the 't~Hooft coupling would be, 
though technically involved, quite straightforward. 
What is necessary for the calculation of the sub-leading terms consists of an asymptotic expansion of a 
given integral depending on a large parameter, and therefore any extra information is needed for the 
calculation. 
We demonstrate that such a calculation is possible for a simple Chern-Simons-matter theory. 
Since our method is based on the planar resolvent, the interpolation between the weak coupling and the strong coupling is 
straightforward, at least qualitatively. 
The interpolation is realized by a continuous change of the parameters, which are functions of the 't Hooft coupling, in the planar resolvent. 

According to our method, we reproduce known results on the large 't~Hooft coupling behavior of the eigenvalue distributions 
and the Wilson loops. 
The theories investigated in this paper are pure Chern-Simons theory, U$(N)_k$ Chern-Simons theory coupled to two adjoint matters, 
ABJM theory and GT theory all of which have at least ${\cal N}=3$ supersymmetry so that the formula in \cite{Kapustin:2009kz} 
is available. 
Here GT-theory refers to one of the Chern-Simons-matter theories constructed in \cite{Gaiotto:2009mv} 
which has ${\cal N}=3$ supersymmetry and has 
the same field content as ABJM theory but the two Chern-Simons levels can be arbitrary. 
The planar limit of this theory was discussed previously in \cite{Suyama:2010hr}. 

\vspace{5mm}

This paper is organized as follows. 
Section \ref{N=3_CSM} contains a brief summary of the theories we discuss in this paper. 
Pure Chern-Simons theory is discussed in section \ref{pureCS} as a warm-up. 
In section \ref{cosech}, we discuss U$(N)_k$ Chern-Simons theory coupled to two adjoint matters. 
This would be the simplest example which exhibits a seemingly 
typical behavior of Chern-Simons-matter theories with a non-vanishing sum of the levels. 
The famous results on ABJM theory are reproduced in section \ref{ABJM} with the method developed in the previous sections. 
In section \ref{GT}, we investigate GT theory in a manner parallel to the ABJM theory. 
It turns out that the structure of the planar resolvents of ABJM theory and GT theory are quite similar, but the behavior in the 
large 't~Hooft coupling limit is qualitatively different. 
In section \ref{sub-leading}, the sub-leading order calculation is presented for the theory discussed in 
section \ref{cosech}. 
Section \ref{discuss} is devoted to discussion. 
Details of the calculations for those theories are summarized in Appendix \ref{details}. 
Appendix \ref{SUSY breaking} discusses the supersymmetry breaking in the context of the matrix model; it looks difficult to 
know the presence of the supersymmetry breaking in Chern-Simons-matter theories from the calculations in the corresponding 
matrix models.

\vspace{1cm}

\section{${\cal N}=3$ Chern-Simons-matter theories and matrix models} \label{N=3_CSM}

\vspace{5mm}

The family of ${\cal N}=3$ Chern-Simons-matter theories is a good arena for the study of various aspects of 
Chern-Simons-matter theories, 
especially in the regime where the perturbative calculation is not available. 
The ${\cal N}=3$ supersymmetry allows one to include any number of matter fields in any representations of the gauge group, 
as long as they form ${\cal N}=4$ hypermultiplets. 
See e.g. \cite{Gaiotto:2007qi}. 
This means that there are a large number of theories of various kinds. 
On the other hand, the ${\cal N}=3$ supersymmetry is powerful enough to constrain quantum corrections, so that the analysis 
of such theories are far easier than the analysis of ${\cal N}\le2$ theories. 
For example, the R-charges of the fundamental fields are not renormalized, and therefore it is rather straightforward to check 
whether a theory is conformal even quantum mechanically \cite{Gaiotto:2007qi}. 

Following \cite{Kapustin:2009kz}, we shall study ${\cal N}=3$ Chern-Simons-matter theories by localization 
technique. 
For a Chern-Simons-matter theory defined on $S^3$ with gauge group $\prod_lU(N_l)_{k_l}$, the partition function is given by 
\begin{equation}
Z \ =\ \int\prod_{l,i_l}du_{l,i_l}\,\exp\left[-S_{\rm tree}-S_{\rm vector}-S_{\rm matter} \right], 
   \label{KWY}
\end{equation}
where 
\begin{eqnarray}
S_{\rm tree} &=& \sum_{l,i_l}\frac{k_l}{4\pi i}(u_{l,i_l})^2, \\
S_{\rm vector} &=& -\sum_{l}\sum_{i_l<j_l}\log\left[ \sinh^2 \frac{u_{l,i_l}-u_{l,j_l}}2 \right], \\
S_{\rm matter} &=& \sum_RS_{{\rm matter},R}, 
\end{eqnarray}
which is applicable to any ${\cal N}=3$ theories. 
The explicit form of $S_{{\rm matter},R}$ depends on the representation $R$ of the corresponding matter field. 
For the fundamental and the adjoint matters for U$(N_l)$ and the bi-fundamental matter for U$(N_l)\times$U$(N_{l'})$, 
it is given as 
\begin{eqnarray}
S_{\rm matter,fund} &=& \sum_{i_l}\log\left[ \cosh\frac{u_{l,i_l}}{2} \right], \\
S_{\rm matter,adj} &=& \sum_{i_l<j_l}\log\left[ \cosh\frac{u_{l,i_l}-u_{l,j_l}}2 \right], \\
S_{\rm matter,bi-fund} &=& \sum_{i_l,j_{l'}}\log\left[ \cosh\frac{u_{l,i_l}-u_{l',j_{l'}}}2 \right]. 
\end{eqnarray}
These formulae were extended to ${\cal N}=2$ theories in \cite{Jafferis:2010un}\cite{Hama:2010av}. 

One remarkable property of the above localization 
formula is that the partition function (\ref{KWY}) looks like that of a multi-matrix model in 
which the angular variables are integrated out. 
In fact, in the case of ABJM theory, the partition function is directly related to a known matrix model \cite{Aganagic:2002wv} 
which was solved exactly \cite{Aganagic:2002wv}\cite{Halmagyi:2003ze}. 
Even for other theories, the simplicity of the integrand in (\ref{KWY}) may allow 
one to deal with them using techniques developed in the context of matrix models. 

Some quantities can be evaluated with the help of this localization formula. 
One such quantity is the free energy which was evaluated for various theories recently 
\cite{Drukker:2010nc}\cite{Herzog:2010hf}\cite{Martelli:2011qj}\cite{Cheon:2011vi}\cite{Jafferis:2011zi}\cite{Niarchos:2011sn}\cite{Gulotta:2011si}\cite{Amariti:2011da}\cite{Amariti:2011xp}. 
Another quantity is the expectation value of BPS Wilson loops \cite{Gaiotto:2007qi}. 
For each unitary factor U$(N_l)_{k_l}$, there is a Wilson loop operator $W_l$ which preserves 1/3 of the 
${\cal N}=3$ supersymmetry. 
The expectation value is given as 
\begin{equation}
\langle W_l \rangle \ =\ \left\langle \frac1{N_l}\sum_{i_l}e^{u_{l,i_l}} \right\rangle_{\rm mm},  
   \label{vev Wilson}
\end{equation}
where the right-hand side is defined as 
\begin{equation}
\langle {\cal O} \rangle_{\rm mm} 
 \ :=\ \frac1Z\int\prod_{l,i_l}du_{l,i_l}\,\exp\left[-S_{\rm tree}-S_{\rm vector}-S_{\rm matter} \right]{\cal O}(u). 
\end{equation}

\vspace{5mm}

It is possible to discuss the 't~Hooft limit, in which the ranks and the levels of every unitary factors U$(N_l)_{k_l}$ are sent 
to infinity, while the ratios 
\begin{equation}
\lambda_l \ :=\ \frac{N_l}{k_l}
\end{equation}
are kept finite. 
In this limit, the saddle-point approximation for the $u$-integrals becomes exact. 
Recall that such a limit can be taken for theories which does not have matter fields in higher representations. 
In fact, the allowed representations are the (anit-)fundamental, bi-fundamental, adjoint and (anti-)symmetric. 
If necessary, it is possible to assign a kind of quiver diagram, including unoriented lines corresponding to (anti-)symmetric 
matters, to each theory which admits the 't~Hooft limit. 
Since the superpotential is determined completely by the supersymmetry, the quiver diagram uniquely specifies the theory. 
Since the 't~Hooft expansion is naturally related to the genus expansion of a dual string theory, 
it may be natural to restrict 
ourselves to such a sub-family of the ${\cal N}=3$ theories as long as our interest is on the relation to Type IIA string theory 
via AdS/CFT correspondence. 
When one would like to take the M-theory limit, according to \cite{Herzog:2010hf}, 
it might be better to consider a broader sub-family of the 
${\cal N}=3$ theories. 

In a suitable limit, (\ref{vev Wilson}) can be written as 
\begin{equation}
\langle W_l \rangle \ =\ \int dx\,\rho_l(x) e^x, 
\end{equation}
where $\rho_l$ is a smooth function of compact support. 
Let $x_*$ be the supremum of the support of $\rho_l$. 
Since $\rho_l$ is defined such that it is positive and its integral is 1, an inequality 
\begin{equation}
\langle W_l \rangle \ \le\ e^{x_*} 
\end{equation}
holds. 
This implies that, if $x_*$ is small, then $\langle W_l \rangle$ cannot be large. 
Therefore, if one would like to have a large value of $\langle W_l \rangle$, which may be expected 
for a large 't~Hooft coupling from AdS/CFT correspondence, 
then $x_*$ must be large. 
In this case, the expectation value of $W_l$ can be estimated as follows 
\begin{equation}
\langle W_l \rangle \ \sim\ e^{x_*}. 
\end{equation}
It will turn out later that the infimum of the support of $\rho_l$ may be negative and large when $x_*$ is large, 
implying that the eigenvalues are 
widely distributed. 
The arguments above is valid even when $u_{l,i_l}$ are complex, and therefore $x_*$ (defined in a suitable manner) is also complex, 
as long as Re$(x_*)$ is large. 

\vspace{5mm}

Although the integrand in (\ref{KWY}) are written in terms of the elementary functions, it is still difficult to solve the 
matrix model exactly, with some exceptions including ABJM theory and pure Chern-Simons theory. 
Perturbative results in terms of the small 't~Hooft couplings can be derived rather easily. 
In fact, one does not need to know the exact planar resolvent for this perturbative calculations. 
This was demonstrated in \cite{Suyama:2009pd} for the evaluation of the Wilson loop in ABJM theory. 

It also turned out that the analysis in the large 't~Hooft coupling regime is indeed tractable. 
As observed in \cite{Suyama:2009pd}, a drastic simplification occurs when the width of the eigenvalue distribution is huge. 
For example, if $|x|$ is large, then one can use 
\begin{equation}
\coth(x) \ =\ \mbox{sgn}(x) + O(e^{-|x|})
\end{equation}
to simplify the saddle-point equations. 
According to the argument above, this approximation is valid for a large 't~Hooft coupling. 
This kind of approximation is available since the complicated one-loop terms are written in terms of the exponential functions. 
A similar approximation is valid for the M-theory limit \cite{Herzog:2010hf}. 
It was efficiently used, at the level of the integrand, to obtain the 
eigenvalue distribution functions in addition to the configurations of the condensed eigenvalues. 
In fact, those results contain all information on operators which preserve the supercharge used in the localization. 
As a result, the analysis of \cite{Herzog:2010hf} nicely derived various results which were 
expected from AdS/CFT correspondence. 

\vspace{5mm}

It turned out \cite{Suyama:2010hr} 
that there are some Chern-Simons-matter theories, in addition to ABJM theory and pure Chern-Simons 
theory, whose planar resolvent can be obtained in terms of contour integrals. 
Although the planar resolvent is not as explicit as in the case for ABJM theory, it is determined up to a few parameters, and 
the equations which determine those parameters were also obtained. 
In the following, we show that a simplification similar to the one mentioned above also occurs in the calculations using the 
integral representation of the planar 
resolvent, so that it is possible to reproduce some exact results obtained so far. 
One typical example of the approximation used in the following is 
\begin{equation}
\sqrt{(e^u-e^{-\alpha})(e^\alpha-e^u)} \ \sim\ \left\{
\begin{array}{cc}
\pm i, & (u_{r}\ll -\alpha_{r}) \\ e^{\frac12(u+\alpha)}, & (-\alpha_{r}\ll u_{r}\ll \alpha_{r}) \\
\pm ie^u, & (\alpha_{r}\ll u_{r})
\end{array}
\right.
\end{equation}
where $u_{r}=\mbox{Re}(u)$ and $\alpha_{r}=\mbox{Re}(\alpha)$. 
Here $\alpha_{r}>0$ is assumed.

\vspace{1cm}

\section{Pure Chern-Simons theory} \label{pureCS}

\vspace{5mm}

${\cal N}=3$ pure Chern-Simons theory is the simplest one among the Chern-Simons-matter theories of interest. 
The matrix model \cite{Marino:2002fk} 
related to this theory has also the simplest structure. 
In fact, it can be solved easily in the large $N$ limit \cite{Aganagic:2002wv}\cite{Halmagyi:2003ze}. 
It is reasonable to expect that analyzing this simple 
matrix model will be very helpful to understand some structures which 
also appear in more complicated matrix models. 

We are interested in the large $N$ limit. 
All the information in this limit can be obtained by solving the saddle-point equations. 
In this section, we consider the following equations \cite{Marino:2002fk}
\begin{equation}
\frac k{2\pi}u_i \ =\  \sum_{j\ne i}\coth\frac{u_i-u_j}2. 
   \label{SP for pureCS}
\end{equation}
The indices $i,j$ run from 1 to $N$. 
The variables $u_i$ are assumed to be real when $k$ is real. 
If $k$ is replaced with $-ik$, then the equations (\ref{SP for pureCS}) 
are the saddle-point equations derived from the localization 
formula (\ref{KWY}) for ${\cal N}=3$ U$(N)_k$ pure Chern-Simons theory. 
Due to this analytic continuation, it will turn out that $u_i$ become imaginary. 
The expectation value $\langle W \rangle$ of the BPS Wilson loop \cite{Gaiotto:2007qi} is 
given as 
\begin{equation}
\langle W \rangle \ =\  \left\langle \frac1N\sum_{i=1}^N e^{u_i} \right\rangle_{\rm mm}.  
   \label{Wilson}
\end{equation}

\vspace{5mm}

It is convenient to introduce new variables $z_i:=e^{u_i}$. 
In terms of them, the equations (\ref{SP for pureCS}) can be written as 
\begin{equation}
\log z_i \ =\ t+2t\cdot \frac1N\sum_{j\ne i}\frac{z_j}{z_i-z_j}
   \label{SP for pureCS2}
\end{equation}
in the large $N$ limit, where 
\begin{equation}
t \ := \ \frac{2\pi N}k
\end{equation}
is the 't~Hooft coupling. 
Following \cite{Aganagic:2002wv}\cite{Halmagyi:2003ze}, the planar resolvent is defined as 
\begin{equation}
v(z) \ :=\  t\int dx\,\rho(x)\frac x{z-x}, 
   \label{def of resolvent}
\end{equation}
where $\rho(x)$ is formally defined as 
\begin{equation}
\rho(x) \ :=\ \frac1N\sum_{i=1}^N\delta(x-z_i). 
\end{equation}
As usual, $\rho(x)$ is assumed to become a smooth function in the large $N$ limit. 

The equation (\ref{SP for pureCS2}) can be interpreted as the requirement of 
the balance between an external force (left-hand side with the opposite sign) 
and a repulsive force between eigenvalues (right-hand side including the constant term $t$). 
Note that the repulsive force is a long-range force which is non-vanishing even for infinitely separated pairs of eigenvalues. 
As long as $t$ is small, the eigenvalue distribution is determined by the external force. 
It is easy to see that the eigenvalues are distributed around $z=1$ at which the external force vanishes. 
Therefore, it is natural to assume that the support of $\rho(x)$ is $[a,b]$ where $0<a<1<b$ for a non-zero $t$. 
The equation (\ref{SP for pureCS2}) then determines the planar resolvent to be 
\begin{equation}
v(z) \ =\  \int_a^b\frac{dx}{2\pi}\frac{\log(e^{-t}x)}{z-x}\frac{\sqrt{(z-a)(z-b)}}{\sqrt{|(x-a)(x-b)|}} 
   \label{CS resolvent}
\end{equation}
for $z\in\mathbb{C}\backslash[a,b]$. 
The explicit expression after performing the contour integration is found in \cite{Aganagic:2002wv}\cite{Halmagyi:2003ze}. 

The definition (\ref{def of resolvent}) of the planar resolvent $v(z)$ implies 
\begin{equation}
v(z) \ =\  \left\{
\begin{array}{cc}
-t, & (z=0) \\ O(z^{-1}). & (z\to\infty) 
\end{array}
\right.
\end{equation}
These two conditions are equivalent if $ab=1$ which is assumed in the following. 
Note that this is expected from the invariance of the equations (\ref{SP for pureCS}) under the simultaneous 
flip of the sign of $u_i$. 
The condition at $z=0$ implies 
\begin{equation}
t \ =\  -\int_a^b\frac{dx}{\pi}\frac{\log x}{x\sqrt{|(x-a)(x-b)|}}. 
   \label{pureCS tHooft}
\end{equation}
This equation determines $a$ in terms of $t$. 

To obtain results for the pure Chern-Simons theory for which $t$ is purely imaginary, one may start with this expression 
for a real $a$, and then analytically continue $a$, while keeping $ab=1$, such that $t$ is purely imaginary. 
Then, the resolvent (\ref{CS resolvent}) with $a$ so determined will have all the information on pure Chern-Simons theory in the 
large $N$ limit. 

In fact, it is not difficult to perform the integration in (\ref{pureCS tHooft}). 
As a result, $t$ and $\alpha:=\log b$ are related as 
\begin{equation}
e^{\frac12\alpha}+e^{-\frac12\alpha} \ =\ 2e^{\frac12t}. 
   \label{CS relation}
\end{equation}
This relation defines a holomorphic map from $\alpha\in\mathbb{C}$ to $t\in\mathbb{C}$. 
This map is a composition of the exponential map and the Zhukovski transformation which is well-known in fluid dynamics. 
If $\alpha\gg 1$, then $t$ behaves as 
\begin{equation}
t(\alpha) \ =\ \alpha-2\log2+O(e^{-\alpha}). 
   \label{approx}
\end{equation}
By rotating the phase, $\alpha\to e^{i\theta}\alpha$, one can make $t$ complex. 
The above expression for $t$ is still valid as long as $|\theta|<\frac\pi2$. 
However, it is not possible to deduce from (\ref{approx}) the relation between $t$ and 
$\alpha$ when $t$ is purely imaginary. 
There is an alternative way to obtain a purely imaginary $t$. 
One may notice from the exact relation (\ref{CS relation}) 
that 
\begin{equation}
t(\alpha+2\pi i) \ =\ t(\alpha) + 2\pi i
    \label{shift}
\end{equation}
holds. 
Therefore, a desired value of purely imaginary $t$ may be obtained by choosing a suitable value of $\alpha\ge0$ 
which is of order one and then shifting $\alpha$ in the imaginary direction. 
This means that it is not enough to just specify the value of $a$ appearing in the resolvent (\ref{CS resolvent}) to obtain 
a large imaginary part of $t$. 

\vspace{5mm}

In the following, we show that the above results can be obtained without invoking the explicit relation (\ref{CS relation}). 
For most of Chern-Simons-matter theories, it seems to be difficult to obtain an explicit expression for the planar resolvent, and 
therefore, an explicit expression for the 't~Hooft coupling like (\ref{CS relation}). 
To analyze such theories, one has to develop a technique which does not rely on any explicit expressions. 
Fortunately, at least integral representations 
like (\ref{CS resolvent})(\ref{pureCS tHooft}) can be obtained for some Chern-Simons-matter theories. 
The following calculations on the pure Chern-Simons matrix model will provide us with some experiences on handling those integral 
representations which will be applied to more complicated theories in the later sections. 

First, we determine the relation between $t$ and $\alpha$, assuming that the length of the branch cut is long. 
Let us focus on the case of real $t$ so that $\alpha$ is also real. 
In terms of a new variable $u:=\log x$, 
\begin{equation}
t \ =\  \int_{-\alpha}^{+\alpha}\frac{du}{\pi}\frac{ue^{\frac{u-\alpha}2}}{\sqrt{(1-e^{-(\alpha+u)})(1-e^{-(\alpha-u)})}}. 
   \label{CS tHooft2}
\end{equation}
One possible approximation valid for large $\alpha$ is to replace the denominator 
\begin{equation}
\sqrt{(1-e^{-(\alpha+u)})(1-e^{-(\alpha-u)})}
\end{equation}
with 1 for most of the range of $u$, as mentioned at the end of section \ref{N=3_CSM}. 
In fact, although this approximation is valid in some cases discussed later, it is not the case here. 
This approximation is valid if most of the values of $u$ contribute equally, while here the dominant contribution is localized 
at $u=\alpha$. 
A crude estimate of the integral (\ref{CS tHooft2}), taking this fact into account, provides 
\begin{equation}
t \ =\ \alpha + O(1). 
   \label{limit relation}
\end{equation}
The details of this estimate is shown in Appendix \ref{estimate_CS}. 

\begin{figure}[tbp]
\begin{minipage}{.5\linewidth}
\includegraphics{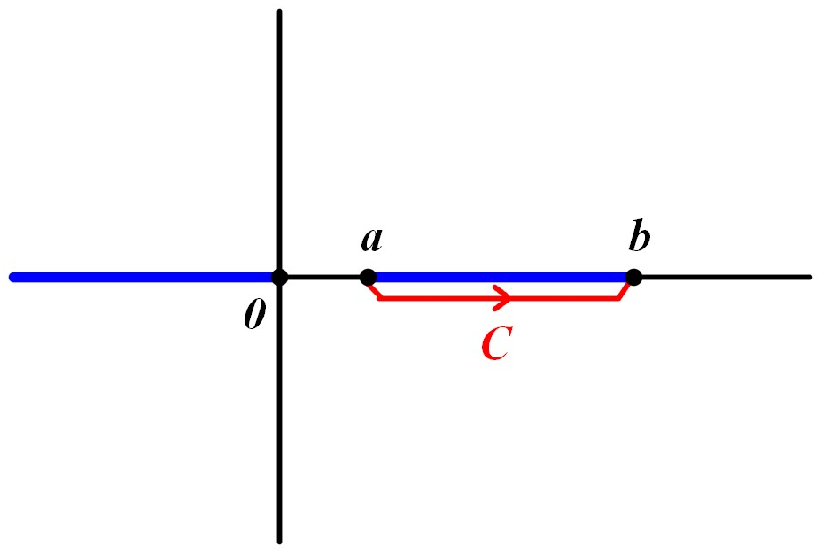}
\caption{The integration contour (red line) and \newline the branch cuts (blue lines) before the shift of $\alpha$. }
   \label{CS1}
\end{minipage}
\begin{minipage}{.45\linewidth}
\includegraphics{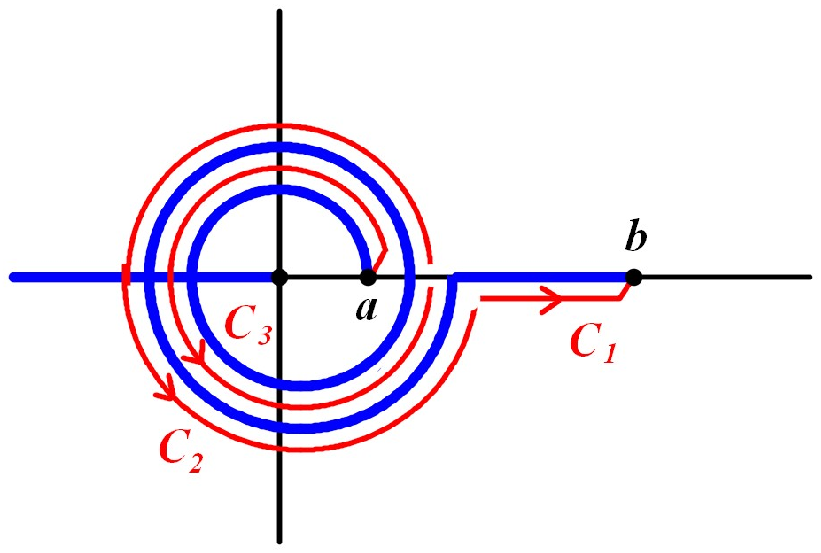}
\caption{The integration contours and the branch cuts after the shift of  by $2\pi i$. }
   \label{CS2}
\end{minipage}
\end{figure}

The property (\ref{shift}) can be also derived from (\ref{pureCS tHooft}). 
To show this, it is convenient to rewrite it as follows, 
\begin{equation}
t \ =\ -\int_C\frac{dx}{\pi i}\frac{\log x}{x\sqrt{(x-a)(x-b)}}, 
\end{equation}
where the contour $C$ runs from $a$ to $b$ below the branch cut, as depicted in Figure \ref{CS1}. 
The shift of $\alpha$ in the imaginary direction rotates the phase of $a$. 
After the shift by $2\pi i$, the contour and the cuts are deformed as depicted in Figure \ref{CS2}. 
The integration along $C_1$ gives 
\begin{eqnarray}
-\int_{C_1}\frac{dx}{\pi i}\frac{\log x}{x\sqrt{(x-a)(x-b)}} 
&=& -\int_{C}\frac{dx}{\pi i}\frac{\log x}{x\sqrt{(x-a)(x-b)}}-\int_{C}\frac{dx}{\pi i}\frac{2\pi i}{x\sqrt{(x-a)(x-b)}} 
    \nonumber \\
&=& t(\alpha)-2\pi i. 
\end{eqnarray}
The integration along $C_3$ gives 
\begin{eqnarray}
-\int_{C_3}\frac{dx}{\pi i}\frac{\log x}{x\sqrt{(x-a)(x-b)}}
&=& \int_{C_2}\frac{dx}{\pi i}\frac{\log x}{x\sqrt{(x-a)(x-b)}}-\int_{C_2}\frac{dx}{\pi i}\frac{2\pi i}{x\sqrt{(x-a)(x-b)}}
    \nonumber \\
&=& \int_{C_2}\frac{dx}{\pi i}\frac{\log x}{x\sqrt{(x-a)(x-b)}}+4\pi i. 
\end{eqnarray}
These calculations show that (\ref{shift}) holds. 

To see the analytic dependence of $t$ on $\alpha$ in general, another expression 
\begin{equation}
t \ =\ -\int_{C'}\frac{dx}{2\pi i}\frac{\log x}{x\sqrt{(x-a)(x-b)}}, 
\end{equation}
where $C'$ encircles the cut $[a,b]$, is useful. 
As long as $a\ne0$, there is a suitable choice of $C'$ such that the integrand is bounded in a neighborhood 
of $C'$ and $a$. 
Therefore, the derivative with respect to $\bar{\alpha}$ commutes with the integral, implying that $t$ is 
analytic in $\alpha$ for a suitable region in $\mathbb{C}$. 
Note that some singularity may appear in the limit $a\to0$ in which the contour $C'$ will be pinched by 
two branch points. 
This corresponds to the divergence of $t$ in the limit $\alpha\to\infty$, as seen above. 
This may be regarded as another indication that a large 't~Hooft coupling is obtained only when the branch cut 
becomes infinitely long. 

The same argument can be applied to more general Chern-Simons-matter theories, as long as the 't~Hooft 
coupling is given in terms of contour integrals. 
In later sections, we understand that the 't~Hooft coupling is an analytic function of parameters. 

\vspace{5mm}

The relation (\ref{limit relation}) implies that, as long as Re$(\alpha)$ is large, the BPS Wilson loop behaves as 
\begin{equation}
\langle W \rangle \sim e^t. 
\end{equation}
As mentioned above, in the Chern-Simons matrix model, 
there exists a long-range repulsive force between eigenvalues which makes the eigenvalue distribution 
wider than that in the matrix models with $1/x$-type interaction. 
A matrix model of the latter kind is the Gaussian matrix model whose saddle-point equations are 
\begin{equation}
u_i \ =\ 2t\cdot\frac 1N\sum_{j\ne i}\frac1{u_i-u_j}. 
\end{equation}
As is well-known in the context of AdS/CFT correspondence, the ``Wilson loop'' defined similarly to 
the right-hand side of (\ref{Wilson}) 
behaves as 
\begin{equation}
\langle W \rangle \sim e^{\sqrt{2t}}. 
\end{equation}
This is a result showing that the presence of a long-range force may change the behavior of the Wilson loop for a large 
't~Hooft coupling in a drastic manner. 

On the other hand, for a purely imaginary $t$, the Wilson loop is not larger than any function of the form $\exp[ct^\gamma]$ 
for any $c,\gamma>0$. 
This is consistent with the exact result on the Wilson loop \cite{Witten:1988hf} in the pure Chern-Simons theory\footnote{
It should be noted that the results of this matrix model for $t=2\pi i\lambda$ with $\lambda>1$ might 
have nothing to do with ${\cal N}=3$ 
pure Chern-Simons theory since the supersymmetry of the pure Chern-Simons theory is spontaneously broken when $\lambda>1$ 
\cite{Kitao:1998mf}\cite{Bergman:1999na}\cite{Ohta:1999iv}, 
and therefore, the relation to the matrix model (\ref{SP for pureCS}) is not obvious. 
}.

\vspace{1cm}

\section{U$(N)_k$ Chern-Simons theory coupled to two adjoints} \label{cosech}

\vspace{5mm}

The next equations we consider are 
\begin{equation}
\frac{k}{2\pi}u_i \ =\ \sum_{j\ne i}\coth\frac{u_i-u_j}{2}-\sum_j\tanh\frac{u_i-u_j}2. 
   \label{SP cosech}
\end{equation}
The indices $i,j$ run from 1 to $N$. 
If $k$ is replaced with $-ik$, then this equations are the same as the saddle-point equations of ${\cal N}=3$ U$(N)_k$ Chern-Simons 
theory coupled to two adjoint matters (i.e. matters including two ${\cal N}=4$ adjoint hypermultiplets). 
The number of matters is chosen such that the long-range repulsive force is absent. 
These equations were analyzed in \cite{Suyama:2009pd} using a technique similar to \cite{Herzog:2010hf}. 

In terms of $z_i=e^{u_i}$, the equations (\ref{SP cosech}) can be written as 
\begin{equation}
\log z_i \ =\ 2t\cdot \frac1N\sum_{j\ne i}\frac{z_j}{z_i-z_j}-2t\cdot \frac1N\sum_j\frac{(-z_j)}{z_i-(-z_j)} 
\end{equation}
in the large $N$ limit. 
This is equivalent to 
\begin{equation}
-\log (-(-z_i)) \ =\ 2t\cdot \frac1N\sum_{j}\frac{z_j}{(-z_i)-z_j}-2t\cdot \frac1N\sum_{j\ne i}\frac{(-z_j)}{(-z_i)-(-z_j)}.  
\end{equation}
These two sets of equations are quite similar to the saddle-point equations of ABJM matrix model \cite{Marino:2009jd}. 
The main difference is that, although the planar resolvent for (\ref{SP cosech}) has two cuts, the positions of 
them are correlated to each other. 

The planar resolvent $v(z)$ of this system is defined as 
\begin{equation}
v(z) \ := \ t\int dx\,\rho(x)\frac x{z-x}-t\int dx\,\rho(-x)\frac x{z-x}. 
\end{equation}
The saddle-point equations determine $v(z)$ to be 
\begin{equation}
v(z) \ =\ 2z\int_a^b\frac{dx}{2\pi}\frac{\log x}{z^2-x^2}\frac{\sqrt{(z^2-a^2)(z^2-b^2)}}{\sqrt{|(x^2-a^2)(x^2-b^2)|}}. 
\end{equation}
The planar resolvent must satisfy the following conditions 
\begin{equation}
v(z) \ =\ \left\{
\begin{array}{cc}
0, & (z=0) \\ O(z^{-1}). & (z\to\infty)
\end{array}
\right.
\end{equation}
The first condition is trivially satisfied. 
The second one implies 
\begin{equation}
\int_a^b\frac{dx}{2\pi}\frac{\log x}{\sqrt{|(x^2-a^2)(x^2-b^2)|}} \ =\ 0. 
\end{equation}
This is satisfied if $ab=1$. 
In the following, we employ this choice. 

To find the relation between $t$ and $\alpha=\log b$, we have to use 
\begin{equation}
t \ =\ \oint_C\frac{dz}{2\pi i}\frac{v(z)}{z}, 
\end{equation}
where the contour $C$ encircles the cut $[a,b]$, since here $v(0)$ is not related to $t$. 
This can be written as 
\begin{eqnarray}
t 
&=& 4\int_a^b\frac{dx}{2\pi}\int_a^b\frac{dx}{2\pi}\log x\frac{\sqrt{|(y^2-a^2)(y^2-b^2)|}}{\sqrt{|(x^2-a^2)(x^2-b^2)|}}
    \,\mbox{P}\frac1{x^2-y^2} \nonumber \\
&=& 2\alpha^3\int_{-1}^{+1}\frac{dv}{2\pi}\int_{-1}^{+1}\frac{du}{2\pi}\,u\frac{\sqrt{(1-e^{-2\alpha(1+v)})
    (1-e^{-2\alpha(1-v)})}}{\sqrt{(1-e^{-2\alpha(1+u)})(1-e^{-2\alpha(1-u)})}}\,\mbox{P}\coth[\alpha(u-v)],  
\end{eqnarray}
where $\alpha u=\log x$ and $\alpha v=\log y$. 
The leading behavior of $t$ for a large $\alpha$ is 
\begin{eqnarray}
t 
&\sim& 2\alpha^3\int_{-1}^{+1}\frac{dv}{2\pi}\int_{-1}^{+1}\frac{du}{2\pi}\,u\,\mbox{sgn}(u-v) \nonumber \\
&=& \frac{2\alpha^3}{3\pi^2}. 
   \label{cosech relation}
\end{eqnarray}
This implies that the Wilson loop behaves as 
\begin{equation}
\log \langle W \rangle \ \sim\ \left( \frac{3\pi^2}{2}t \right)^\frac13. 
\end{equation}
This coincides with the result obtained in \cite{Suyama:2009pd}. 
The behavior of $\langle W \rangle$ for a large 't~Hooft coupling is different from the one observed in ${\cal N}=4$ super 
Yang-Mills theory and ABJM theory, but still it is exponentially increasing with the 't~Hooft coupling. 
A crude estimate shows that sub-leading terms are smaller than $\alpha^{2+\epsilon}$ for any $\epsilon>0$. 
Probably, the sub-leading term is of order $O(\alpha^2)$. 

\begin{figure}[tbp]
\begin{minipage}{.53\linewidth}
\includegraphics{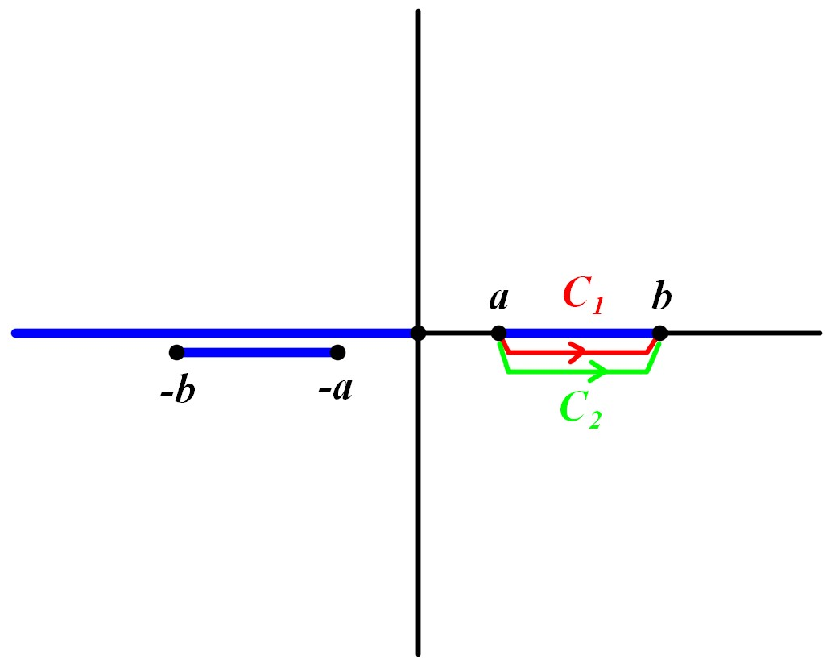}
\caption{The integration contour (red and \newline green lines) and the branch cuts (blue lines) \newline 
         before the shift of $\alpha$. }
   \label{cosech1}
\end{minipage}
\begin{minipage}{.45\linewidth}
\vspace*{-4mm}
\includegraphics{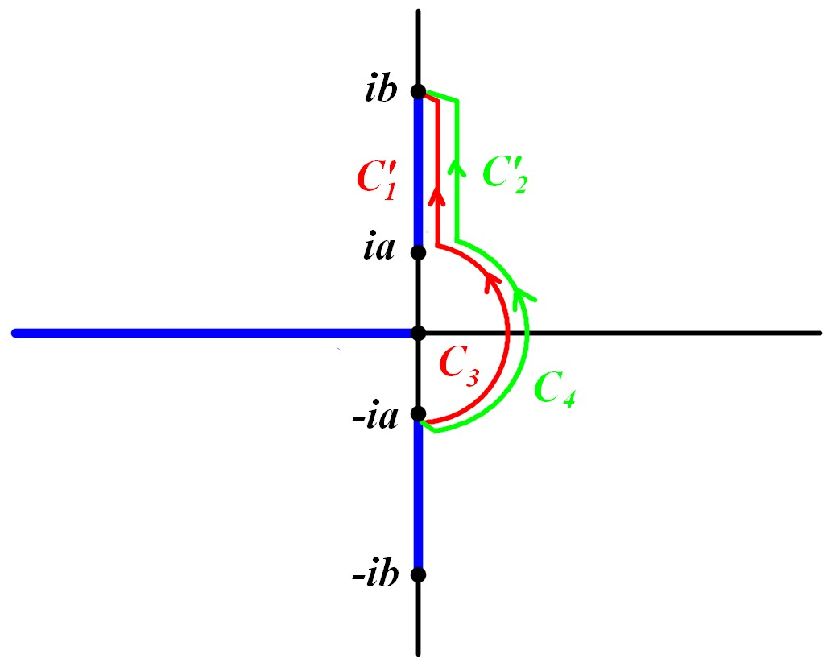}
\caption{The integration contours and the branch cuts after the shift of $\alpha$ by $\frac{\pi i}2$. }
   \label{cosech2}
\end{minipage}
\end{figure}

The analytic continuation to a purely imaginary $t$ is now straightforward. 
It can be done by rotating the phase of $\alpha$ by $\frac\pi6$. 
During the rotation, Re$(\alpha)$ is kept large, and therefore, the above estimate of $t$ is still valid. 
It is also possible to perform the analytic continuation as was done for the pure Chern-Simons matrix model. 
The 't~Hooft coupling $t$ can be written as 
\begin{equation}
t \ =\ 4\int_{C_1}\frac{dy}{2\pi i}\int_{C_2}\frac{dx}{2\pi i}\frac{\log x}{y^2-x^2}\frac{\sqrt{(y^2-a^2)(y^2-b^2)}}
       {\sqrt{(x^2-a^2)(x^2-b^2)}}, 
\end{equation}
where the contours $C_1,C_2$ are depicted in Figure \ref{cosech1}. 
A shift of $\alpha$ by $\frac\pi2i$ deforms the cuts and contours as in Figure \ref{cosech2}. 
For large $\alpha$, the shifted quantity $t(\alpha+\frac\pi2i)$ is estimated to be 
\begin{equation}
t\left( \alpha+\frac{\pi} 2i \right)-t(\alpha) \ \sim\ \frac i\pi\alpha^2. 
   \label{ac cosech}
\end{equation}
Details of the estimate are shown in Appendix \ref{estimate cosech}. 
This is consistent with (\ref{cosech relation}). 

\vspace{5mm}

It is interesting to estimate the free energy of the theory. 
The free energy is given as 
\begin{equation}
F \ \sim\ \left\langle k\sum_{i=1}^Nu_i^2 + (\mbox{one-loop}) \right\rangle_{\rm mm}. 
\end{equation}
Assuming that the Gaussian part and the one-loop part provide contributions of the same order, it scales as 
\begin{equation}
F \ \sim \ k^\frac13N^\frac53 
\end{equation}
up to an overall constant independent of $N$ and $k$. 
This is the same scaling found in \cite{Jafferis:2011zi} in the M-theory limit. 
Since the M-theory limit in \cite{Jafferis:2011zi} 
is not the same as the large $N$ limit here, it might not be necessary for those results to match. 
However, it was shown \cite{Drukker:2010nc}\cite{Herzog:2010hf} that, 
in the case of ABJM theory, the functional forms of the free energy in the M-theory limit and 
the 't~Hooft limit coincide with each other. 
It may be expected that the behavior of the eigenvalue distribution found above would have some general nature for theories whose 
Chern-Simons levels does not sum to zero. 
Later, it will be shown that the same behavior do appear in GT theory. 

\vspace{5mm}

It may be interesting to see the effect of a long-range force by comparing the above results with those of 
\begin{equation}
\frac{k}{2\pi}u_i \ =\ \sum_{j\ne i}\coth\frac{u_i-u_j}2+\sum_{j}\tanh\frac{u_i-u_j}2. 
   \label{modified}
\end{equation}
In this system, a long-range force exists. 
In fact, these equations can be written as 
\begin{equation}
\frac{k}{4\pi}u_i \ =\ \sum_{j\ne i}\coth(u_i-u_j). 
\end{equation}
This is equivalent to the saddle-point equations (\ref{SP for pureCS}) of pure Chern-Simons theory. 
As was shown in section \ref{pureCS}, 
the behavior of the eigenvalue distribution of pure Chern-Simons theory is quite different from 
that of the Chern-Simons-matter theory discussed in this section. 
In particular, the solution of the equations (\ref{modified}) does not have a long branch cut for a purely imaginary $t$, 
even when its absolute value is large.

\vspace{1cm}

\section{ABJM theory} \label{ABJM}

\vspace{5mm}

The third example is ABJM theory. 
The goal of this section is to derive the famous result \cite{Marino:2009jd} 
\begin{equation}
\langle W \rangle \ \sim \ e^{\pi\sqrt{2\lambda}}, \hspace{5mm} (\lambda\ \to\ +\infty)
   \label{ABJM behavior}
\end{equation}
where $\lambda:=\frac Nk$, from the integral representation of the planar resolvent \cite{Suyama:2010hr}. 

\vspace{5mm}

Recall the saddle-point equations of ABJ theory \cite{Aharony:2008gk}. 
\begin{eqnarray}
\frac{k}{2\pi i}u_i 
&=& \sum_{j\ne i}\coth\frac{u_i-u_j}2-\sum_{a}\tanh\frac{u_i-v_a}2, \\
-\frac{k}{2\pi i}v_a 
&=& \sum_{b\ne a}\coth\frac{v_a-v_b}2-\sum_i\tanh\frac{v_a-u_i}2. 
\end{eqnarray}
The indices $i,j$ run from 1 to $N_1$ and $a,b$ from 1 to $N_2$. 
Notice that there are two sets of eigenvalues, $\{u_i\}$ and $\{v_a\}$, according to the two gauge group factors 
U$(N_1)_k\times$U$(N_2)_{-k}$. 
Introducing new variables $z_i:=e^{u_i}$ and $w_a:=e^{v_a}$, the above equations can be written as 
\begin{eqnarray}
\log z_i 
&=& t_1-t_2+2t_1\cdot \frac1{N_1}\sum_{j\ne i}\frac{z_j}{z_i-z_j}-2t_2\cdot\frac1{N_2}\sum_a\frac{(-w_a)}{z_i-(-w_a)}, \\
\log (-(-w_a))
&=& t_1-t_2+2t_1\cdot\frac1{N_1}\sum_i\frac{z_i}{(-w_a)-z_i}-2t_2\cdot\frac1{N_2}\sum_{b\ne a}\frac{(-w_a)}{(-w_a)-(-w_b)}, 
\end{eqnarray}
where 
\begin{equation}
t_1 \ :=\ \frac{2\pi iN_1}{k}, \hspace{5mm} t_2 \ :=\ \frac{2\pi iN_2}{k}.   
\end{equation}
In the following, $t_1$ and $t_2$ are regarded as real variables, and they will be analytically continued to the above values 
later. 
The structure of the saddle-point equations suggests that there would be two cuts, one is around $z=+1$ where $z_i$ condense, 
and the other is around $z=-1$ where $-w_a$ condense. 
Define the planar resolvent 
\begin{equation}
v(z) \ := \ t_1\int dx\,\rho(x)\frac x{z-x}-t_2\int dx\,\tilde{\rho}(x)\frac x{z-x}. 
   \label{ABJM resolvent}
\end{equation}
The distribution of $z_i$ is described by $\rho(x)$, and that of $-w_a$ is described by $\tilde{\rho}(x)$. 
We assume that the supports of $\tilde{\rho}(x)$ and $\rho(x)$ are $[a,b]$ and $[c,d]$, respectively, where $b<0<c$.  
The saddle-point equations determine $v(z)$ to be 
\begin{eqnarray}
v(z) 
&=& -\int_a^b\frac{dx}{2\pi}\frac{\log(-e^{t_2-t_1}x)}{z-x}\frac{\sqrt{(z-a)(z-b)(z-c)(z-d)}}{\sqrt{|(x-a)(x-b)(x-c)(x-d)|}} 
    \nonumber \\
& & +\int_c^d\frac{dx}{2\pi}\frac{\log (e^{t_2-t_1}x)}{z-x}\frac{\sqrt{(z-a)(z-b)(z-c)(z-d)}}{\sqrt{|(x-a)(x-b)(x-c)(x-d)|}}. 
   \label{ABJM explicit resolvent}
\end{eqnarray}
The conditions imposed on $v(z)$ are 
\begin{equation}
v(x) \ =\ \left\{ 
\begin{array}{cc}
t_2-t_1, & (z=0), \\ O(z^{-1}). & (z\to\infty)
\end{array}
\right.
\end{equation}
This amounts to three conditions on the parameters $a,b,c,d$. 
These are reduced to one condition, assuming 
\begin{equation}
ab \ =\ 1, \hspace{5mm} cd \ =\ 1. 
\end{equation}
The remaining equation 
\begin{eqnarray}
\frac{t_2-t_1}2 
&=& -\int_a^b\frac{dx}{2\pi}\frac{\log (-x)}{x\sqrt{|(x-a)(x-b)(x-c)(x-d)|}} \nonumber \\
& & +\int_c^d\frac{dx}{2\pi}\frac{\log x}{x\sqrt{|(x-a)(x-b)(x-c)(x-d)|}}. 
   \label{ABJM tHooft}
\end{eqnarray}
provides a relation among the undetermined parameters and the 't~Hooft couplings. 
One more condition is necessary to completely determine the parameters. 
One may use one of the following relations 
\begin{equation}
t_1 \ =\ \oint_{C_+}\frac{dz}{2\pi i}\frac{v(z)}z, \hspace{5mm} t_2 \ =\ -\oint_{C_-}\frac{dz}{2\pi i}\frac{v(z)}z,  
    \label{cycles}
\end{equation}
where $C_+$ encircles the cut $[c,d]$ and $C_-$ encircles $[a,b]$. 
It was shown in \cite{Suyama:2010hr} that the relation (\ref{ABJM tHooft}) can be written explicitly as follows, 
\begin{equation}
(c+d)-(a+b) \ =\ 4e^{t_1-t_2}. 
    \label{ABJM relation}
\end{equation}

\vspace{5mm}

First, let us focus on the case $t_1=t_2$, that is, the ranks of two gauge groups are equal, $N_1=N_2=N$. 
Suppose that both $|a|$ and $d$ are large for which the corresponding 't~Hooft couplings are large. 
Then, the relation (\ref{ABJM relation}) implies 
\begin{equation}
a+b \ =\ \Lambda-2, \hspace{5mm} c+d \ =\ \Lambda+2, 
   \label{approx relation}
\end{equation}
where $|\Lambda|$ is large. 
This relation can be satisfied only after an analytic continuation, since originally $a<0<d$ was assumed. 
The resulting branch cuts should be almost parallel to each other, but the direction of the cuts in the complex plane is not 
determined by (\ref{approx relation}). 

Here, recalling the exact solution found in \cite{Marino:2009jd} 
will be helpful to find a more detailed information of the branch cuts. 
It was shown that 
\begin{equation}
\alpha \ =\ 2+i\kappa, \hspace{5mm} \beta \ =\ 2-i\kappa 
\end{equation}
holds, where $\alpha=c+d$ and $\beta=-a-b$ in our notation. 
Here $\kappa$ is a real function of the 't~Hooft coupling which is large and positive when the 't~Hooft coupling is large. 
This indicates that the lengths of the cuts are huge as expected above, and 
the cuts are almost parallel to the imaginary axis. 

With this input, we show that the integral representation of the planar resolvent (\ref{ABJM explicit resolvent}) 
can reproduce the relation between the 't~Hooft 
coupling and the length of the branch cuts. 
We assume 
\begin{equation}
a \ =\ -ie^\gamma-2+O(e^{-\gamma}), \hspace{5mm} d \ =\ -ie^\gamma+2+O(e^{-\gamma}), 
   \label{ABJM ansatz}
\end{equation}
where $\gamma>0$ is assumed to be large. 
Of course, this is compatible with (\ref{approx relation}). 
It can be shown that the flip of signs in the imaginary part of $a,d$ results in the flip of the sign of the 't~Hooft coupling. 

\begin{figure}[tbp]
\begin{minipage}{.5\linewidth}
\includegraphics{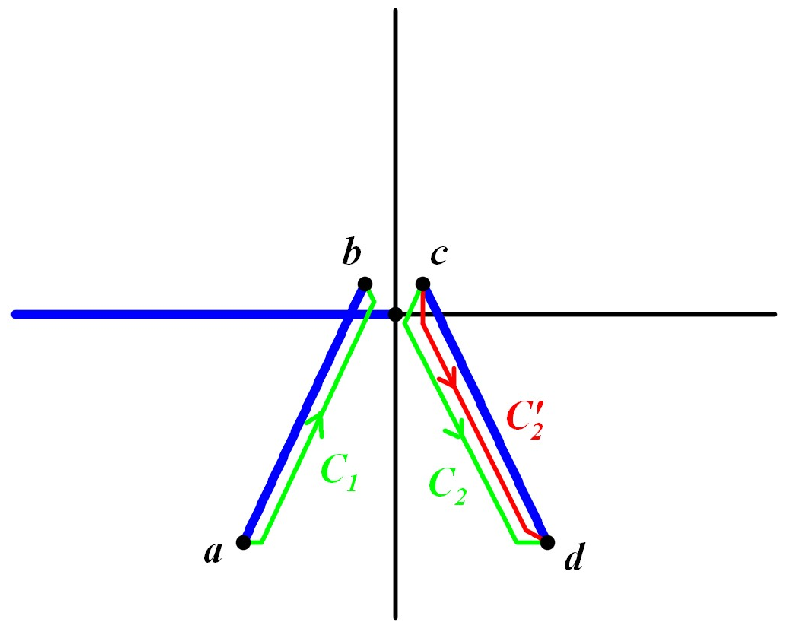}
\caption{The integration contours (red and \newline green lines) and the branch cuts (blue lines) \newline 
         before the shift of $\alpha$. }
   \label{ABJM1}
\end{minipage}
\begin{minipage}{.45\linewidth}
\vspace*{-5mm}
\includegraphics{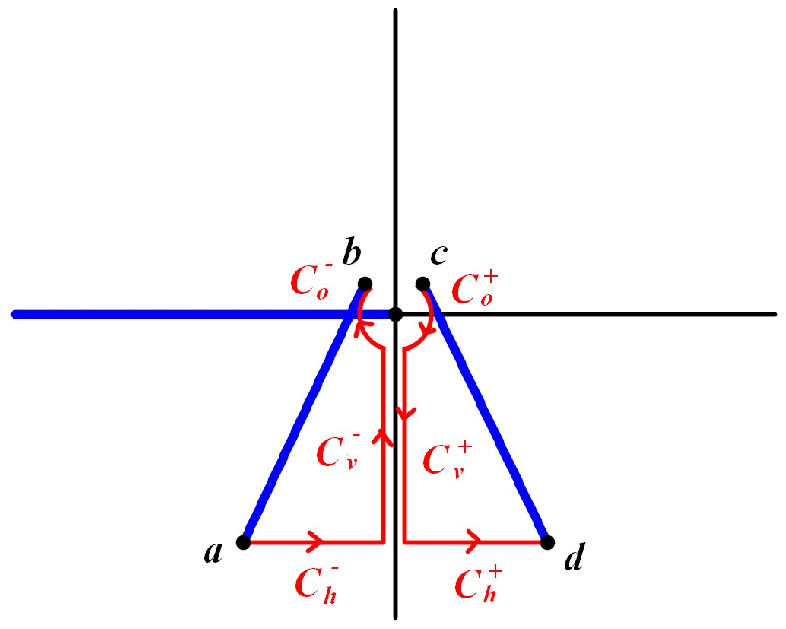}
\caption{The deformed integration contours convenient for estimation. }
   \label{ABJM2}
\end{minipage}
\end{figure}

The 't~Hooft coupling $t_1$ can be written as 
\begin{eqnarray}
t_1 
&=& 2\int_{C_2'}\frac{dy}{2\pi i}\int_{C_1+C_2}\frac{dx}{2\pi i}\frac{\log x}{y(y-x)}\frac{\sqrt{(y-a)(y-b)(y-c)(y-d)}}
    {\sqrt{(x-a)(x-b)(x-c)(x-d)}} \nonumber \\
& & +2\pi i\int_{C_2'}\frac{dy}{2\pi i}\int_{C_1}\frac{dx}{2\pi i}\frac1{y(y-x)}\frac{\sqrt{(y-a)(y-b)(y-c)(y-d)}}
    {\sqrt{(x-a)(x-b)(x-c)(x-d)}}. 
\end{eqnarray}
The contours are depicted in Figure \ref{ABJM1}. 
To estimate the integrals, it is convenient to deform the contour as in Figure \ref{ABJM2}. 
It turns out that the first term in the right-hand side is of order $O(\gamma)$. 
The dominant contribution coming from the second term is 
\begin{equation}
2\pi i\int_{C_v^+}\frac{dy}{2\pi i}\int_{C_v^-}\frac{dx}{2\pi i}\frac1{y(y-x)}\frac{\sqrt{(y-a)(y-b)(y-c)(y-d)}}
    {\sqrt{(x-a)(x-b)(x-c)(x-d)}}. 
\end{equation}
The other terms are negligible compared to this term. 
By estimating this integral, one obtains the following asymptotic behavior 
\begin{equation}
t_1 \ \sim\  \frac i\pi\gamma^2. 
   \label{ABJM asymptotic}
\end{equation}
See Appendix \ref{estimate ABJM} for the details. 
Recall that $t_1 = 2\pi i\lambda$ and $\lambda=\frac Nk$. 
Then $\gamma$ can be solved in terms of $\lambda$ as  
\begin{equation}
\gamma \ \sim\  \pi\sqrt{2\lambda}. 
\end{equation}
This then implies the desired behavior (\ref{ABJM behavior}) of the Wilson loop for large $\lambda$ 
including the exact coefficient in the exponent. 

In retrospect, it would be possible to find the right configuration of the branch cuts without relying on the exact 
solution in \cite{Marino:2009jd}. 
One may start with the ansatz (\ref{ABJM ansatz}) which can be deduced from (\ref{ABJM relation}). 
One may try with a real $\gamma$, and obtain the result (\ref{ABJM asymptotic}). 
Recall that (\ref{ABJM relation}) allows a complex $\gamma$. 
Since $t$ should be purely imaginary, one cannot rotate the phase of $\gamma$. 
It is still possible to shift $\gamma$ in the imaginary direction by an $O(1)$ amount. 
This does not change (\ref{ABJM asymptotic}), but does change the $O(\gamma)$ term by a real amount. 
Therefore, whether such a shift is necessary or not can be determined by estimating the $O(\gamma)$ terms. 
This then determines the direction of the cuts. 
Anyway, if one is only interested in the leading order behavior, the direction of the cuts turned out to be irrelevant. 

\vspace{5mm}

The possibility of an interpolation between the weak coupling region and the strong coupling region for, say, the Wilson loop is obvious. 
As was shown in \cite{Suyama:2009pd} that the small 't~Hooft coupling limit corresponds to a limit in which the branch cuts 
of the planar resolvent shrink to a point. 
It is easy to imagine a continuous deformation of the branch cuts from the point-like one to the one depicted in Figure 
\ref{ABJM1}, while keeping the conditions $ab=cd=1$. 
This indicates that the weak coupling results and the strong coupling results can be connected continuously.

\vspace{5mm}

Next, let us consider the case $t_1\ne t_2$ and both $t_1$ and $t_2$ are purely imaginary. 
Recall the relation 
\begin{equation}
(c+d)-(a+b) \ =\ 4e^{t_1-t_2}. 
\end{equation}
The right-hand side is of order one, and therefore, the positions of the cuts are the same as those of ABJM theory 
at the leading order of the large 't~Hooft couplings. 
A natural guess for the sub-leading terms would be  
\begin{equation}
a \ \sim\ -ie^{\gamma}-2e^{t_1-t_2}, \hspace{5mm} d \ \sim \ -ie^\gamma+2e^{t_1-t_2}. 
\end{equation}
It can be shown that the resulting relation between $\gamma$ and $t_1$ are the same at the leading order of $\gamma$. 
As a result, the Wilson loops in ABJ theory behaves as 
\begin{equation}
\langle W \rangle \ \sim \ \exp\left[ \pi\sqrt{\lambda_1+\lambda_2} \right], \hspace{5mm} 
 \lambda_{1,2} \ :=\  \frac{N_{1,2}}{k}. 
\end{equation}
This was obtained in \cite{Marino:2009jd}. 

It was claimed in \cite{Aharony:2008gk} that ABJ theory is well-defined only for $|N_1-N_2|\le k$. 
This implies 
\begin{equation}
|t_1-t_2|\le 2\pi 
   \label{ABJ region}
\end{equation}
is allowed to consider. 
From the matrix model point of view, however, it seems to be difficult to find any 
sign of the ill-definedness for the corresponding 
parameter region. 
Indeed, an analytic continuation by which (\ref{ABJ region}) is violated 
does not show any singular behavior in the planar resolvent, nor in the Wilson 
loop. 
It is similar in pure Chern-Simons theory where the matrix model is well-defined for any value of the 't~Hooft coupling. 
In addition, the Wilson loop does not show any special behavior at $\lambda=1$. 
In the appendix \ref{SUSY breaking}, 
the analysis on ${\cal N}=3$ U$(N)_k$ Chern-Simons theory coupled to fundamental matters is summarized. 
In this theory, the supersymmetry is broken for a choice of parameters which is expected 
from the argument based on the brane construction. 
However, there is also no special behavior of the Wilson loop at a value of the 't Hooft coupling at which 
the supersymmetry breaking is expected to occur.

\vspace{1cm}

\section{GT theory} \label{GT}

\vspace{5mm}

One of the straightforward generalization of ABJM theory, at least from the point of view of saddle-point equations, is 
GT theory \cite{Gaiotto:2009mv}. 
This is proposed to be dual to a massive Type IIA string theory\cite{Gaiotto:2009mv}\cite{Fujita:2009kw}. 
The localization procedure of \cite{Kapustin:2009kz} can be applied to GT theory. 
The resulting saddle-point equations are 
\begin{eqnarray}
\frac{k_1}{2\pi i}u_i 
&=& \sum_{j\ne i}\coth\frac{u_i-u_j}2-\sum_{a}\tanh\frac{u_i-v_a}2, \\
\frac{k_2}{2\pi i}v_a 
&=& \sum_{b\ne a}\coth\frac{v_a-v_b}2-\sum_i\tanh\frac{v_a-u_i}2. 
\end{eqnarray}
When $k_1+k_2=0$, GT theory is reduced to ABJM theory. 
The planar resolvent defined as in (\ref{ABJM resolvent}) was obtained in \cite{Suyama:2010hr} 
in terms of the following contour integrals 
\begin{eqnarray}
v(z) 
&=& \kappa_2\int_a^b\frac{dx}{2\pi}\frac{\log(-e^{-\frac1{\kappa_2}(t_2-t_1)}x)}{z-x}\frac{\sqrt{(z-a)(z-b)(z-c)(z-d)}}
    {\sqrt{|(x-a)(x-b)(x-c)(x-d)|}} \nonumber \\
& & +\kappa_1\int_c^d\frac{dx}{2\pi}\frac{\log (e^{\frac1{\kappa_1}(t_2-t_1)}x)}{z-x}\frac{\sqrt{(z-a)(z-b)(z-c)(z-d)}}
    {\sqrt{|(x-a)(x-b)(x-c)(x-d)|}}, 
\end{eqnarray}
where 
\begin{equation}
\kappa_{1,2} \ := \ \frac{k_{1,2}}{k}, \hspace{5mm} t_{1,2} \ := \ \frac{2\pi iN_{1,2}}{k}. 
\end{equation}
All the parameters $k_{1,2}, N_{1,2}$ are assumed to be proportional to a common number $k$ which is sent to infinity. 
As in the case of ABJM theory, assuming $ab=1$ and $cd=1$, the number of the 
parameters in the planar resolvent can be reduced to two, and the remaining parameters 
are then related to the 't~Hooft couplings $t_{1,2}$ as (\ref{cycles}). 
Instead, one of these relations can be replaced with 
\begin{eqnarray}
\frac{t_2-t_1}2 
&=& \kappa_2\int_a^b\frac{dx}{2\pi}\frac{\log (-x)}{x\sqrt{|(x-a)(x-b)(x-c)(x-d)|}} \nonumber \\
& & +\kappa_1\int_c^d\frac{dx}{2\pi}\frac{\log x}{x\sqrt{|(x-a)(x-b)(x-c)(x-d)|}}. 
    \label{GT tHooft}
\end{eqnarray}
Let us consider the case where $\kappa_1+\kappa_2\ne0$ and $t_1=t_2$. 
In this case, an explicit expression for (\ref{GT tHooft}) like (\ref{ABJM relation}) is not known. 
Therefore, it is necessary to extract some information on the branch cuts directly from this integral representation. 

We assume that $\kappa_{1}$ and $\kappa_2$ are of order one. 
Let $a=-e^{\alpha}$ and $d=e^{\beta}$, and both $\alpha$ and $\beta$ are large. 
It turns out that an assumption $\beta\gg \alpha$ 
is not compatible with (\ref{GT tHooft}) since in this case the right-hand side is 
estimated to be $\frac\beta2$. 
It is of course similar for the other assumption $\alpha\gg \beta$. 
Therefore, it should be assumed that $\alpha$ and $\beta$ are of the same order. 
This assumption then implies 
\begin{equation}
\kappa_2 e^{\alpha}-\kappa_1 e^\beta \ \sim\  0. 
   \label{approx GT tHooft}
\end{equation}
This relation also holds for the ABJM case $\kappa_1=-\kappa_2=1$. 
Indeed, in this case there is one solution 
\begin{equation}
\alpha \ \sim \ \gamma+\frac\pi 2i, \hspace{5mm} \beta \ \sim \ \gamma -\frac\pi2i, 
\end{equation}
where $\gamma$ is a large and positive number for which 
\begin{equation}
a \ \sim \ -ie^\gamma, \hspace{5mm} d \ \sim\ -ie^\gamma. 
   \label{approx GT solution}
\end{equation}
This is the ansatz used in the previous section. 
Since $\kappa_1$ and $\kappa_2$ are of order one, the solution of (\ref{approx GT tHooft}) may not be different so much from the 
above one. 
That is, the solution should be the same with (\ref{approx GT solution}) up to an order one correction. 
Note that the solution (\ref{approx GT solution}) is also valid for a complex $\gamma$ as long as Re$(\gamma)$ is large. 

\vspace{5mm}

Now it is possible to estimate $t_1$ as a function of $\gamma$ in the large $\gamma$ limit. 
We assume first that $\gamma$ is real, and its phase will be rotated if necessary. 
The formula for $t_1$ is 
\begin{eqnarray}
t_1 
&=& -2\kappa_2\int_{C_2'}\frac{dy}{2\pi i}\int_{C_1}\frac{dx}{2\pi i}\frac{\log x}{y(y-x)}\frac{\sqrt{(y-a)(y-b)(y-c)(y-d)}}
    {\sqrt{(x-a)(x-b)(x-c)(x-d)}} \nonumber \\
& & +2\kappa_1\int_{C_2'}\frac{dy}{2\pi i}\int_{C_2}\frac{dx}{2\pi i}\frac{\log x}{y(y-x)}\frac{\sqrt{(y-a)(y-b)(y-c)(y-d)}}
    {\sqrt{(x-a)(x-b)(x-c)(x-d)}} \nonumber \\
& & -2\pi i\kappa_2\int_{C_2'}\frac{dy}{2\pi i}\int_{C_1}\frac{dx}{2\pi i}\frac{1}{y(y-x)}\frac{\sqrt{(y-a)(y-b)(y-c)(y-d)}}
    {\sqrt{(x-a)(x-b)(x-c)(x-d)}}. 
   \label{GT tHooft1}
\end{eqnarray}
Since the configuration of the branch cuts are almost the same with that for ABJM theory, the estimate of the integrals proceeds 
in a similar way. 
Especially, one may immediately find that the third integral provides an $O(\gamma^2)$ contribution. 

The difference from the ABJM theory comes from the other two terms. 
One can show that the leading contribution to each integral is of order $O(\gamma^3)$. 
It is larger than the third integral because of the presence of $\log x$ in the integrand which is of order 
$O(\gamma)$ in most of the integration region of $x$. 
What happened in ABJM theory is that this leading contributions exactly cancel between the first and the second terms, 
so that the second largest term 
coming from the third integral determines the asymptotic relation between $t_1$ and $\gamma$. 
Here in GT theory, such a cancellation is not complete, and the remaining $O(\gamma^3)$ term dominates. 
The resulting relation is 
\begin{equation}
t_1 \ \sim\ \frac{\kappa_1+\kappa_2}{3\pi^2}\gamma^3. 
   \label{GT asymptotic result}
\end{equation}
See Appendix \ref{estimate GT} for the details. 
The right-hand side is real when $\gamma$ is real. 
To obtain the purely imaginary $t_1$, one may rotate the phase of $\gamma$ by $\frac\pi6$. 
During this rotation of $\gamma$, Re$(\gamma)$ is kept large, and therefore the above relation is valid. 
In terms of $\lambda:=\frac{N_1}k$, $\gamma$ is given as 
\begin{equation}
\gamma \ =\ |\gamma|e^{\frac{\pi}6i}, \hspace{5mm} |\gamma| \ \sim\ \left( \frac{6\pi^3}{\kappa_1+\kappa_2}\lambda \right)^\frac13. 
   \label{GT result}
\end{equation}
Note that in (\ref{GT result}) 
the number $k$ cancels out, so $\gamma$ is independent of the choice of $k$, as it should be. 

This result is qualitatively similar to the result obtained in section \ref{cosech}. 
As discussed there, this scaling suggests that the free energy of GT theory scales like $k^\frac13N^\frac53$ which coincides with  
the one found in \cite{Jafferis:2011zi} although the limit taken there is not the usual 't~Hooft limit considered here. 
Interestingly, not only the scaling but also the numerical factor is the same with the result in \cite{Jafferis:2011zi}. 
In \cite{Jafferis:2011zi}, the eigenvalues lie on the line 
\begin{equation}
y \ = \ \frac1{\sqrt{3}}x, 
\end{equation}
which corresponds to the analytic continuation of $\gamma$ mentioned above. 
The maximum value $x_*$ of $x$ is determined by $\rho(x_*)=0$ where 
\begin{equation}
\rho(x) \ =\ \frac{3^\frac16k^\frac13}{2\pi p^\frac13}-\frac{2kx^2}{3\sqrt{3}\pi^3p}, 
\end{equation}
where $k:=k_1+k_2$. 
The necklace quiver with $p=2$ in \cite{Jafferis:2011zi} is GT theory. 
One obtains 
\begin{equation}
x_* \ =\ \frac{\sqrt{3}}{2}\left( \frac{6\pi^3}k \right)^\frac13
\end{equation}
which coincides with the result obtained above. 

Another immediate consequence of (\ref{GT result}) is that the Wilson loop of GT theory behaves as 
\begin{equation}
|\langle W \rangle| \ \sim \ \exp\left[ \frac{\sqrt{3}}2\left( \frac{6\pi^3}{\kappa_1+\kappa_2}\lambda \right)^\frac13 \right]. 
   \label{GT WL}
\end{equation}
Since $N_1=N_2$ was assumed in the above calculations, this formula is valid for Wilson loops for both U$(N)$ factors. 
This is actually consistent\footnote{
We would like to thank Soo-Jong Rey for drawing our attention to this issue, and checking independently 
that the scaling (\ref{GT WL}) agrees with supergravity. 
} with the results in \cite{Aharony:2010af} 
where it was argued that the radius $l$ of the dual 
AdS$_4$ in massive Type IIA supergravity is 
\begin{equation}
l \ \sim\ \frac{N^\frac16}{n_0^\frac16}, 
   \label{massive AdS}
\end{equation}
where $n_0$ is proportional to the $F_0$ flux. 
Since the exponent of $|\langle W \rangle|$ should be proportional to an area in the AdS$_4$, the Wilson loop should scale 
like (\ref{GT WL}). 
Note that the scaling (\ref{massive AdS}) was shown to be valid if 
\begin{equation}
N \ \gg\ \frac{k^3}{n_0^2}, 
\end{equation}
which is indeed satisfied in our setup.

\vspace{1cm}

\section{Sub-leading contributions} \label{sub-leading}

\vspace{5mm}

\begin{figure}[htbp]
\hspace*{4cm}
\includegraphics{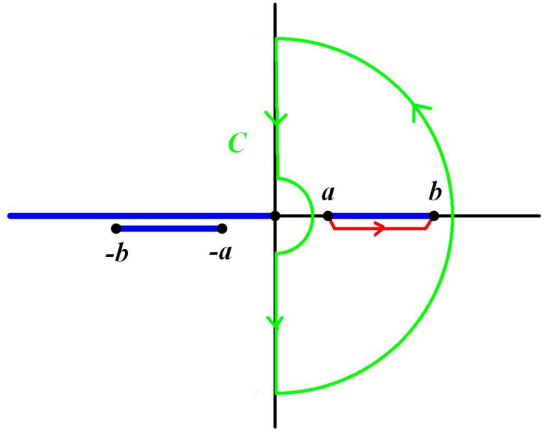}
\caption{The integration contour (red and green lines) and the branch cuts (blue lines).}
   \label{cosech3}
\end{figure}

In this section, we present the calculation of the sub-leading contributions to $t(\alpha)$ for the theory 
considered in section \ref{cosech}. 
To do this, it turns out that the following expression for $t(\alpha)$ is convenient. 
\begin{equation}
t(\alpha) \ =\ 2\int_C\frac{dy}{2\pi i}\int_{-\alpha}^{+\alpha}\frac{du}{2\pi}\frac{u\,e^{-\alpha}}
 {y^2-e^{2u}}\frac{\sqrt{(y^2-a^2)(y^2-b^2)}}{\sqrt{(1-e^{-2(\alpha+u)})(1-e^{-2(\alpha-u)})}}, 
\end{equation}
where the contour $C$ is depicted in Figure \ref{cosech3}. 

The dominant contributions come from the following parts of the integral 
\begin{eqnarray}
& & 2\left[ \int_{ib}^{ia}+\int_{-ia}^{-ib} \right]\frac{dy}{2\pi i}\int_{-\alpha}^{+\alpha}
    \frac{du}{2\pi}\frac{u\,e^{-\alpha}}{y^2-e^{2u}}\frac{\sqrt{(y^2-a^2)(y^2-b^2)}}
    {\sqrt{(1-e^{-2(\alpha+u)})(1-e^{-2(\alpha-u)})}} \nonumber \\
&=& \frac1{2\pi^2}\int_{-\alpha}^{+\alpha}dv\int_{-\alpha}^{+\alpha}du\,u\,\mbox{sgn}(u-v)\left[ 
    1-\frac{e^{-|u-v|}}{\cosh(u-v)}  \right]\frac{\sqrt{(1+e^{-2(\alpha+v)})(1+e^{-2(\alpha-v)})}}
    {\sqrt{(1-e^{-2(\alpha+u)})(1-e^{-2(\alpha-u)})}}. 
\end{eqnarray}
This integral is divided into two parts. 
The first part is evaluated as follows, 
\begin{eqnarray}
& & \frac1{2\pi^2}\int_{-\alpha}^{+\alpha}dv\int_{-\alpha}^{+\alpha}du\,u\,\mbox{sgn}(u-v)
    \frac{\sqrt{(1+e^{-2(\alpha+v)})(1+e^{-2(\alpha-v)})}}
    {\sqrt{(1-e^{-2(\alpha+u)})(1-e^{-2(\alpha-u)})}} \nonumber \\
&=& \frac{2}{\pi^2}\int_0^{\alpha}du\int_0^udv\,u\frac{\sqrt{(1+e^{-2(\alpha+v)})(1+e^{-2(\alpha-v)})}}
    {\sqrt{(1-e^{-2(\alpha+u)})(1-e^{-2(\alpha-u)})}} \nonumber \\
&\sim& \frac{2}{\pi^2}\int_0^{\alpha}du\int_0^udv\,u\frac{\sqrt{1+e^{-2(\alpha-v})}}
    {\sqrt{1-e^{-2(\alpha-u)}}} \nonumber \\
\end{eqnarray}
where exponentially small terms are neglected. 
Now $v$-integration can be performed exactly. 
As a result, one obtains
\begin{eqnarray}
& & \frac2{\pi^2}\int_0^{\alpha}du\frac u{\sqrt{1-e^{-2(\alpha-u)}}}\left[ u+\sqrt{1+e^{-2(\alpha-u)}} 
    -1+\log2-\log(1+\sqrt{1+e^{-2(\alpha-u)}})  \right] \nonumber \\
&=& \frac2{3\pi^2}\alpha^3+O(\alpha), 
\end{eqnarray}
This coincides with (\ref{cosech relation}). 
Note that the coefficients of $O(\alpha)$ terms can be determined. 
For example, an $O(\alpha)$ integral can be estimated as follows, 
\begin{eqnarray}
& & \frac2{\pi^2}\int_0^{\alpha}du\frac u{\sqrt{1-e^{-2(\alpha-u)}}}(\sqrt{1+e^{-2(\alpha-u)}}-1)
    \nonumber \\
&=& \frac2{\pi^2}\int_0^{\alpha}dt\frac{(\alpha-t)}{\sqrt{1-e^{-2t}}}(\sqrt{1+e^{-2t}}-1) \nonumber \\
&=& \frac2{\pi^2}\alpha\int_0^{\infty}dt\frac{1}{\sqrt{1-e^{-2t}}}(\sqrt{1+e^{-2t}}-1)+O(1). 
\end{eqnarray}
The last equality holds 
since the integrals in the second line are convergent when the integration region $[0,\alpha]$ is replaced 
with $[0,\infty]$. 

The second part turns out to be of order $O(\alpha)$. 
Indeed, 
\begin{eqnarray}
& & -\frac1{2\pi^2}\int_{-\alpha}^{+\alpha}dv\int_{-\alpha}^{+\alpha}du\,u\,\mbox{sgn}(u-v)
    \frac{e^{-|u-v|}}{\cosh(u-v)} \frac{\sqrt{(1+e^{-2(\alpha+v)})(1+e^{-2(\alpha-v)})}}
    {\sqrt{(1-e^{-2(\alpha+u)})(1-e^{-2(\alpha-u)})}} \nonumber \\
&\sim& -\frac4{\pi^2}\alpha\int_0^{\infty}d\xi\frac{\xi e^{-2\xi}}{\cosh(2\xi)}. 
\end{eqnarray}

\vspace{5mm}

The contributions from the other parts of the contour $C$ turn out to be of order $O(\alpha)$. 
For example, 
\begin{eqnarray}
& & 2\int_{C_+}\frac{dy}{2\pi i}\int_{-\alpha}^{+\alpha}\frac{du}{2\pi}\frac{u\,e^{-\alpha}}
    {y^2-e^{2u}}\frac{\sqrt{(y^2-a^2)(y^2-b^2)}}{\sqrt{(1-e^{-2(\alpha+u)})(1-e^{-2(\alpha-u)})}} 
    \nonumber \\
&=& \frac1{2\pi^2}\int_{-\frac \pi2}^{+\frac \pi2}d\theta\int_{-\alpha}^{+\alpha}du
    \frac{ue^{\epsilon+i\theta}e^{2(u-\alpha)}}{e^{2\epsilon+2i\theta}-e^{2(u-\alpha)}}
    \frac{\sqrt{(1-e^{-4\alpha-2\epsilon-2i\theta})(1-e^{-2\epsilon-2i\theta})}}
    {\sqrt{(1-e^{-2(\alpha+u)})(1-e^{-2(\alpha-u)})}} \nonumber \\
&\sim& \frac1{2\pi^2}\int_{-\frac \pi2}^{+\frac \pi2}d\theta\int_{0}^{+\alpha}du
    \frac{ue^{\epsilon+i\theta}e^{2(u-\alpha)}}{e^{2\epsilon+2i\theta}-e^{2(u-\alpha)}}
    \frac{\sqrt{(1-e^{-4\alpha-2\epsilon-2i\theta})(1-e^{-2\epsilon-2i\theta})}}
    {\sqrt{1-e^{-2(\alpha-u)}}} \nonumber \\
&=& \frac1{2\pi^2}\alpha\int_{-\frac \pi2}^{+\frac \pi2}d\theta\int_{0}^{\infty}dt
    \frac{e^{\epsilon+i\theta}e^{-2t}}{e^{2\epsilon+2i\theta}-e^{-2t}}
    \frac{\sqrt{1-e^{-2\epsilon-2i\theta}}}
    {\sqrt{1-e^{-2t}}}+O(1). 
\end{eqnarray}

In summary, the asymptotic expansion of $t(\alpha)$ for large $\alpha$ is 
\begin{equation}
t(\alpha) \ =\ \frac{2}{3\pi^2}\alpha^3+c_1\alpha+O(1), 
\end{equation}
where $c_1$ is a constant which can be calculated at least numerically. 
Inverting the relation, one obtains 
\begin{equation}
\alpha(t) \ = \ \left( \frac{3\pi^2}{2} \right)^\frac13t^\frac13
 -\frac13\left( \frac{3\pi^2}{2} \right)^\frac23t^{-\frac13}+O(t^{-\frac23}). 
\end{equation}

\vspace{1cm}

\section{Discussion} \label{discuss}

\vspace{5mm}

We have shown that some exact planar results on the eigenvalue distributions in matrix models related to Chern-Simons-matter theories 
can be derived from integral representations of the resolvents. 
All the results derived in this paper were already derived using a topological string theory \cite{Marino:2009jd} 
or a saddle-point approximation for the 
localization formula of the partition function \cite{Herzog:2010hf}, but the method in this paper is new. 
Our method requires the determination of the planar resolvent, but it does not need to be as explicit as 
in the case of ABJM theory. 
In addition, the method to derived the information on the eigenvalue distribution is a rather elementary estimate of an 
ordinary single- or double integrals. 
The systematic expansion around the large 't~Hooft coupling limit is possible without any 
conceptual difficulty. 
Since the resolvent is determined for any value of the 't~Hooft couplings, up to the exact positions of the branch points, the 
existence of an interpolating function of, for example, the expectation value of the BPS Wilson loop $\langle W \rangle$ is 
evident. 

One of the results obtained in this paper is the relation between 
the 't~Hooft coupling $t$ and a parameter $\alpha$ which determines 
the position of a branch point. 
In general, the relation would be of the form, 
\begin{equation}
t \ = \ c_3\alpha^3+ic_2\alpha^2+c_1\alpha + \cdots, 
\end{equation}
provided that Re$(\alpha)>0$ is large. 
The coefficients $c_{1,2,3}$ are real. 
In general, $c_3$ is non-zero, and then $t$ behaves as $\alpha^3$ even after an analytic continuation which brings $t$ to be 
purely imaginary. 
This seems to be typical when the sum of the Chern-Simons levels are non-zero. 
Such a behavior turns out to be consistent with the conjecture \cite{Gaiotto:2009mv}. 
For a suitable choice of the theory with a suitable choice of the parameters, $c_3=0$ may be realized while $c_2\ne0$. 
Then the theory exhibits some properties in the large 't~Hooft coupling limit which is expected from an AdS$_4$ geometry in an 
ordinary (massless) gravity dual. 
In addition, there exists a special kind of theories for which $c_3=c_2=0$. 
In such theories, the properties in the large 't~Hooft coupling limit may be quite different from the other two kinds of theories 
mentioned above. 
It is interesting that our analysis based on the planar resolvent provides us with a rather unified picture on the structure of 
Chern-Simons-matter theories in the large 't~Hooft coupling limit. 

It seems that our method can be applied to more general theories. 
Our method would be useful since it does not require one to obtain the planar resolvent in the most explicit form. 
In addition, it seems that our method requires less numbers of assumptions compared with \cite{Herzog:2010hf}. 

It would be interesting to apply our method to a family of circular quiver Chern-Simons-matter theories studied recently in 
\cite{Herzog:2010hf}. 
In their research, it was found that the eigenvalue distribution is not always linear but it can be piecewise linear. 
It seems to be difficult to understand this fact from the viewpoint of the planar resolvent since, usually, the eigenvalue 
distribution corresponds to the branch cuts of the resolvent, and the cuts are usually taken to be linear. 
Of course, the configuration of the branch cut is not determined a priori, and therefore, the branch cuts can be chosen to be 
piecewise linear. 
However, if it is the case, then it should be possible to know the reason why it should be so. 
It is also possible that the piecewise-linearity might suggest that the method based on the resolvent does not work for those 
theories. 

Assuming that our method could be applied to a large family of theories, it would be interesting to classify the behavior of the 
eigenvalue distribution in the large 't~Hooft coupling limit. 
In this paper, we found three patterns, two of which have geometric interpretation via AdS/CFT correspondence. 
It would be very surprising if these three exhaust all possible behaviors. 
If it is the case, one might interpret this as an indication that 
the presence of a gravity dual is rather common among ${\cal N}=3$ Chern-Simons-matter theories with a sensible 't~Hooft limit. 
If there exist other behaviors, then it would be interesting to investigate the implication of such behaviors, possibly related 
to a new example of AdS/CFT correspondence. 

The main task for the calculations in this paper is to obtain the asymptotic expansion of a function $t(\alpha)$ which is defined 
in terms of a double integrals with the integrand including $\log x$. 
It would be interesting if there exists a mathematical framework dealing with such functions, enabling one to make 
our calculations transparent.

\vspace{2cm}

{\bf \Large Acknowledgements}

\vspace{5mm}

We would like to thank Soo-Jong Rey for valuable discussions. 
This work was supported by the BK21 program of the Ministry of Education, Science and Technology, 
National Science Foundation of Korea Grants 0429-20100161, R01-2008-000-10656-0, 
2005-084-C00003, 2009-008-0372 and EU-FP Marie Curie Research 
\& Training Network HPRN-CT-2006-035863 (2009-06318).

\appendix

\vspace{1cm}

\section{Estimate of integrals} \label{details}

\vspace{5mm}

\subsection{Pure Chern-Simons theory} \label{estimate_CS}

\vspace{5mm}

We evaluate the following integral 
\begin{equation}
\int_{-\alpha}^{+\alpha}\frac{du}{\pi}\frac{u\,e^{\frac12(u-\alpha)}}{\sqrt{(1-e^{-(\alpha+u)})(1-e^{-(\alpha-u)})}}. 
\end{equation}
This is divided into three parts $I_1+I_2+I_3$ where 
\begin{eqnarray}
I_1 &=& \int_0^{\alpha}\frac{du}{\pi}\frac{\alpha\, e^{\frac12(u-\alpha)}}{\sqrt{(1-e^{-(\alpha+u)})(1-e^{-(\alpha-u)})}}, \\
I_2 &=& \int_0^{\alpha}\frac{du}{\pi}\frac{(u-\alpha) e^{\frac12(u-\alpha)}}{\sqrt{(1-e^{-(\alpha+u)})(1-e^{-(\alpha-u)})}}, \\
I_3 &=& \int_{-\alpha}^{0}\frac{du}{\pi}\frac{u\, e^{\frac12(u-\alpha)}}{\sqrt{(1-e^{-(\alpha+u)})(1-e^{-(\alpha-u)})}}. 
\end{eqnarray}
Noticing 
\begin{equation}
\frac1{\sqrt{1-e^{-2\alpha}}}\ \le \ \frac1{\sqrt{1-e^{-(\alpha+u)}}} \ \le\ \frac1{\sqrt{1-e^{-\alpha}}}
\end{equation}
for $u\in[0,\alpha]$, $I_1$ satisfies  
\begin{equation}
\frac{\alpha}{\sqrt{1-e^{-2\alpha}}}\left( 1-\frac2\pi\sin^{-1}e^{-\frac\alpha2} \right) \ \le I_1 \ \le 
 \frac{\alpha}{\sqrt{1-e^{-\alpha}}}\left( 1-\frac2\pi\sin^{-1}e^{-\frac\alpha2} \right), 
\end{equation}
which implies 
\begin{equation}
I_1 \ =\ \alpha + O(\alpha e^{-\frac\alpha2}). 
\end{equation}
$I_2$ can be estimated as follows. 
\begin{eqnarray}
|I_2| 
&\le& \frac1{\sqrt{1-e^{-\alpha}}}\int_0^\alpha\frac{du}\pi\frac{(\alpha-u)e^{\frac12(u-\alpha)}}{\sqrt{1-e^{u-\alpha}}} 
      \nonumber \\
&\le& \frac1{\sqrt{1-e^{-\alpha}}}\int_0^\infty\frac{dt}{\pi}\frac{t\,e^{-\frac t2}}{\sqrt{1-e^{-t}}}. 
\end{eqnarray}
Therefore, 
\begin{equation}
I_2 \ =\ O(1). 
\end{equation}
The remaining part $I_3$ can be estimated as follows. 
\begin{eqnarray}
|I_3| 
&=& \int_0^\alpha\frac{du}{\pi}\frac{u\,e^{-\frac12(u+\alpha)}}{\sqrt{(1-e^{-(\alpha+u)})(1-e^{-(\alpha-u)})}} \nonumber \\
&\le& \frac{\alpha\,e^{-\frac\alpha2}}{\sqrt{1-e^{-\alpha}}}\left( \frac\alpha\pi+\frac2\pi\log(1+\sqrt{1-e^{-\alpha}}) \right). 
\end{eqnarray}
Therefore, this part is negligible. 
Combining these estimates, we obtained (\ref{limit relation}).

\vspace{5mm}

\subsection{Chern-Simons theory with two adjoints} \label{estimate cosech}

\vspace{5mm}

In section \ref{cosech}, an analytic continuation of 
\begin{equation}
t(\alpha) \ =\ 4\int_{C_1}\frac{dy}{2\pi i}\int_{C_2}\frac{dx}{2\pi i}\frac{\log x}{y^2-x^2}\frac{\sqrt{(y^2-a^2)(y^2-b^2)}}
 {\sqrt{(x^2-a^2)(x^2-b^2)}}  
\end{equation}
is discussed, where $b=e^\alpha$ and $ab=1$. 
Shifting $\alpha$ by $\frac{\pi i}2$, this becomes 
\begin{equation}
t\left( \alpha+\frac{\pi}2i \right) 
 \ =\ 4\int_{C_1'+C_3}\frac{dy}{2\pi i}\int_{C_2'+C_4}\frac{dx}{2\pi i}\frac{\log x}{y^2-x^2}\frac{\sqrt{(y^2+a^2)(y^2+b^2)}}
 {\sqrt{(x^2+a^2)(x^2+b^2)}},  
\end{equation}
where the integration contours are depicted in Figure \ref{cosech2}. 
To estimate the resulting integral, it is convenient to divide it into four integrals labeled by contours. 

The first integral we estimate is the one for the contours $(C_1',C_2')$. 
This can be written as 
\begin{eqnarray}
& & 4\int_{C_1'}\frac{dy}{2\pi i}\int_{C_2'}\frac{dx}{2\pi i}\frac{\log x}{y^2-x^2}\frac{\sqrt{(y^2+a^2)(y^2+b^2)}}
    {\sqrt{(x^2+a^2)(x^2+b^2)}} \nonumber \\
&=& 4\int_{C_1}\frac{dy}{2\pi i}\int_{C_2}\frac{dx}{2\pi i}\frac{\log (ix)}{y^2-x^2}\frac{\sqrt{(y^2-a^2)(y^2-b^2)}}
    {\sqrt{(x^2-a^2)(x^2-b^2)}} \nonumber \\
&=& t(\alpha) + 2\pi i\int_{C_1}\frac{dy}{2\pi i}\int_{C_2}\frac{dx}{2\pi i}\frac{1}{y^2-x^2}\frac{\sqrt{(y^2-a^2)(y^2-b^2)}}
    {\sqrt{(x^2-a^2)(x^2-b^2)}}. 
\end{eqnarray}
The second term of the last line is estimated as follows. 
\begin{eqnarray}
& & -2\pi i\int_a^b\frac{dy}{2\pi}\int_a^b\frac{dx}{2\pi}\frac1{y+x}\left[ \mbox{P}\frac{1}{y-x}-\pi i\delta(y-x) \right]
    \frac{\sqrt{(y^2-a^2)(y^2-b^2)}}{\sqrt{(x^2-a^2)(x^2-b^2)}} \nonumber \\
&=& -2\pi i\int_{-\alpha}^{+\alpha}\frac{dv}{2\pi}\int_{-\alpha}^{+\alpha}\frac{du}{2\pi}\frac{\sqrt{(1-e^{-2(\alpha+v)})
    (1-e^{-2(\alpha-v)})}}{\sqrt{(1-e^{-2(\alpha+u)})(1-e^{-2(\alpha-u)})}}\mbox{P}\frac{e^{v-u}}{e^{v-u}-e^{u-v}}
    +O(\alpha) \nonumber \\
&=& -\pi i\int_{-\alpha}^{+\alpha}\frac{dv}{2\pi}\int_{-\alpha}^{+\alpha}\frac{du}{2\pi}\frac{\sqrt{(1-e^{-2(\alpha-v)})
    (1-e^{-2(\alpha-v)})}}{\sqrt{(1-e^{-2(\alpha+u)})(1-e^{-2(\alpha-u)})}} +O(\alpha) \nonumber \\
&\sim& -\pi i\int_{-\alpha}^{+\alpha}\frac{dv}{2\pi}\int_{-\alpha}^{+\alpha}\frac{du}{2\pi} \nonumber \\
&=& -\frac i{\pi}\alpha^2. 
\end{eqnarray}

Next, consider the contours $(C_1',C_4)$. 
The corresponding integral can be estimated as 
\begin{eqnarray}
& & 4\int_{C_1'}\frac{dy}{2\pi i}\int_{C_4}\frac{dx}{2\pi i}\frac{\log x}{y^2-x^2}\frac{\sqrt{(y^2+a^2)(y^2+b^2)}}
    {\sqrt{(x^2+a^2)(x^2+b^2)}} \nonumber \\
&\sim& 4\int_{C_1'}\frac{dy}{2\pi i}\frac{\sqrt{(y^2+a^2)(y^2+b^2)}}{y^2}\int_{-\frac\pi2}^{+\frac\pi 2}\frac{d\phi}{2\pi}
    \frac{-\alpha\,e^{-\alpha+i\phi}}{\sqrt{e^{2i\phi}+1}} \nonumber \\
&=& 2i\alpha e^{-\alpha}\int_a^b\frac{dy}{2\pi}\frac{\sqrt{(y^2-a^2)(b^2-y^2)}}{y^2} \nonumber \\
&\sim& \frac{2i}{\pi}\alpha^2. 
\end{eqnarray}
It turns out that the remaining integrals are negligible compared to the above terms. 
Therefore, we obtained the estimate (\ref{ac cosech}).

\vspace{5mm}

\subsection{ABJM theory} \label{estimate ABJM}

\vspace{5mm}

It turns out that the dominant contribution to the 't~Hooft coupling $t_1$ comes from the following integral 
\begin{equation}
2\pi i\int_{C_v^+}\frac{dy}{2\pi i}\int_{C_v^-}\frac{dx}{2\pi i}\frac1{y(y-x)}\frac{\sqrt{(y-a)(y-b)(y-c)(y-d)}}
 {\sqrt{(x-a)(x-b)(x-c)(x-d)}}, 
\end{equation}
where 
\begin{equation}
a \ =\ -ie^\gamma-2+O(e^{-\gamma}), \hspace{5mm} d \ =\ -ie^\gamma+2+O(e^{-\gamma}), \hspace{5mm} ab \ =\ cd \ =\ 1, 
\end{equation}
and the integration contours are depicted in Figure \ref{ABJM2}. 
This can be written as follows. 
\begin{eqnarray}
& & 2\pi i\int_{-ie^{-\gamma}+0}^{-ie^{+\gamma}+0}\frac{dy}{2\pi i}\int_{-ie^{+\gamma}-0}^{-ie^{-\gamma}-0}\frac{dx}{2\pi i}
    \frac1{y(y-x)}\frac{\sqrt{(y-a)(y-b)(y-c)(y-d)}}{\sqrt{(x-a)(x-b)(x-c)(x-d)}} \nonumber \\
&=& 2\pi i\int_{e^{-\gamma}+i0}^{e^{+\gamma}+i0}\frac{dy}{2\pi i}\int_{e^{+\gamma}-i0}^{e^{-\gamma}-i0}\frac{dx}{2\pi i}
    \frac1{y(y-x)}\frac{\sqrt{(y-ia)(y-ib)(y-ic)(y-id)}}{\sqrt{(x-ia)(x-ib)(x-ic)(x-id)}} \nonumber \\
&=& 2\pi i\int_{e^{-\gamma}}^{e^{+\gamma}}\frac{dy}{2\pi i}\int_{e^{+\gamma}}^{e^{-\gamma}}\frac{dx}{2\pi i}
    \frac1{y}\left[ \mbox{P}\frac1{y-x}-\pi i\delta(y-x) \right]
    \frac{\sqrt{(y-ia)(y-ib)(y-ic)(y-id)}}{\sqrt{(x-ia)(x-ib)(x-ic)(x-id)}}. 
\end{eqnarray}
The term with the delta function provides a contribution of order $O(\gamma)$. 
The remaining term can be estimated as follows. 
\begin{eqnarray}
& & 2\pi i\int_{e^{-\gamma}}^{e^{+\gamma}}\frac{dy}{2\pi i}\int_{e^{+\gamma}}^{e^{-\gamma}}\frac{dx}{2\pi i}
    \frac1{y}\mbox{P}\frac1{y-x}\frac{\sqrt{(y-ia)(y-ib)(y-ic)(y-id)}}{\sqrt{(x-ia)(x-ib)(x-ic)(x-id)}} \nonumber \\
&\sim& -2\pi i\int_{-\gamma}^{+\gamma}\frac{dv}{2\pi i}\int_{-\gamma}^{+\gamma}\frac{du}{2\pi i}\mbox{P}\frac{e^v}{e^v-e^u}
    \nonumber \\
&=& -\pi i\int_{-\gamma}^{+\gamma}\frac{dv}{2\pi i}\int_{-\gamma}^{+\gamma}\frac{du}{2\pi i} \nonumber \\
&=& \frac i{\pi}\gamma^2. 
\end{eqnarray}

\vspace{5mm}

\subsection{GT theory} \label{estimate GT}

\vspace{5mm}

The dominant contribution to $t_1$ in (\ref{GT tHooft1}) comes from the following two integrals 
\begin{eqnarray}
& & -2\kappa_2\int_{C_v^+}\frac{dy}{2\pi i}\int_{C_v^-}\frac{dx}{2\pi i}\frac{\log x}{y(y-x)}
    \frac{\sqrt{(y-a)(y-b)(y-c)(y-d)}}{\sqrt{(x-a)(x-b)(x-c)(x-d)}} \nonumber \\
& & +2\kappa_1\int_{C_v^+}\frac{dy}{2\pi i}\int_{C_v^+}\frac{dx}{2\pi i}\frac{\log x}{y(y-x)}
    \frac{\sqrt{(y-a)(y-b)(y-c)(y-d)}}{\sqrt{(x-a)(x-b)(x-c)(x-d)}}. 
   \label{main GT}
\end{eqnarray}
Note that they cancel each other if $\kappa_1=\kappa_2$. 
In general, it is expected that these terms provides the contributions of order $O(\gamma^3)$; the length of the integration 
range in terms of $\log x$ and $\log y$ is of order $O(\gamma)$ and $\log x$ in the integrand 
provides another $O(\gamma)$ contribution. 
This $O(\gamma^3)$ contribution 
vanishes only for the ABJM slice, so that the deformation from ABJM theory to GT theory is not continuous in the large 
't~Hooft coupling limit. 

Each integral can be estimated similarly. 
Let us focus on the first one. 
By the similar calculation in the ABJM case, it can be estimated as 
\begin{eqnarray}
& & -2\kappa_2\int_{C_v^+}\frac{dy}{2\pi i}\int_{C_v^-}\frac{dx}{2\pi i}\frac{\log x}{y(y-x)}
    \frac{\sqrt{(y-a)(y-b)(y-c)(y-d)}}{\sqrt{(x-a)(x-b)(x-c)(x-d)}} \nonumber \\
&\sim& 2\kappa_2\int_{-\gamma}^{+\gamma}\frac{dv}{2\pi i}\int_{-\gamma}^{+\gamma}\frac{du}{2\pi i}\,\mbox{P}\frac{u\,e^v}{e^v-e^u}
    \nonumber \\
&=& \kappa_2\int_{-\gamma}^{+\gamma}\frac{dv}{2\pi i}\int_{-\gamma}^{+\gamma}\frac{du}{2\pi i}\,u\,\mbox{P}\coth\frac{u-v}2 
    \nonumber \\
&\sim& \kappa_2\gamma^3\int_{-1}^{+1}\frac{dv}{2\pi i}\int_{-1}^{+1}\frac{du}{2\pi i}\,u\,\mbox{sgn}(u-v) \nonumber \\
&=& \frac{\kappa_2}{3\pi^2}\gamma^3. 
\end{eqnarray}
Note that an extra sign appears in the second integral in (\ref{main GT}) due to the opposite direction of the integration 
contour for $x$. 
As a result, the 't~Hooft coupling $t_1$ behaves as (\ref{GT asymptotic result}) in the large $\gamma$ limit.

\vspace{1cm}

\section{On SUSY breaking in Chern-Simons-flavor theory} \label{SUSY breaking}

\vspace{5mm}

Consider ${\cal N}=2$ pure Chern-Simons theory. 
The expectation value of the normalized BPS Wilson loop is \cite{Kapustin:2009kz} 
\begin{equation}
\langle W \rangle \ \propto\  \frac1N\frac{\sin(\pi N/k)}{\sin(\pi/k)}
\end{equation}
up to an overall phase. 

It is easy to see that $\langle W \rangle$ is non-zero for $0\le \frac Nk<1$, and it vanishes at $\frac Nk=1$. 
Curiously, $\frac Nk=1$ is the boundary of the parameter region in which the supersymmetry is spontaneously broken 
\cite{Kitao:1998mf}\cite{Bergman:1999na}\cite{Ohta:1999iv}. 
Note that this behavior of the Wilson loop is preserved in the 't~Hooft limit. 
Indeed, 
\begin{equation}
\lim_{N\to\infty}\frac1N\frac{\sin(\pi N/k)}{\sin(\pi/k)} 
 \ =\  \lim_{N\to\infty}\frac1N\frac{\sin(\pi\lambda)}{\sin(\pi\lambda/N)}
 \ =\  \frac{\sin(\pi\lambda)}{\pi\lambda},  
   \label{CSWL}
\end{equation}
which vanishes at $\lambda=1$ but is non-zero for $0\le \lambda<1$, where $\lambda=\frac Nk$. 

This might suggest that the vanishing of the Wilson loop might be related to the breaking of supersymmetry. 
To check whether this could be the case, we consider another example, ${\cal N}=3$ Chern-Simons theory coupled to $N_f$ fundamental 
hypermultiplets. 

\vspace{5mm}

First, let us consider the expected pattern of the supersymmetry breaking in ${\cal N}=3$ Chern-Simons theory with flavors. 
We assume that the supersymmetry 
breaking pattern of this theory would be the same as ${\cal N}=2$ Yang-Mills Chern-Simons theory with 
the same gauge group coupled to the same number of fundamental hypermultiplets. 
(We expect that the deformation from ${\cal N}=2$ theory to ${\cal N}=3$ theory, which is a continuous change in the 
superpotential, does not change the vacuum structure.) 
Then, the supersymmetry breaking pattern can be deduced from the corresponding D-brane configuration. 

The brane configuration in question consists of two parallel NS5-branes, $N$ D3-branes suspended between the NS5-branes, $k$ 
D5-branes which are intersecting with one of the NS5-branes, and $N_f$ D5-branes which are parallel to the other set of D5-branes 
and are intersecting with the D3-branes. 
Since the D3-branes have a finite extent in one direction, the worldvolume theory is a three-dimensional one. 
In the absence of the second set of D5-branes, the worldvolume theory on the D3-branes is ${\cal N}=2$ U$(N)_k$ 
Yang-Mills Chern-Simons 
theory, and $N_f$ D5-branes add massless fundamental hypermultiplets to the theory. 
The question of whether the ground state of the theory preserves the supersymmetry can be answered using the s-rule 
\cite{Hanany:1996ie}. 
Roughly speaking, this rule says that only a single D3-brane can be suspended between an NS5-brane and a D5-brane in the above 
setup. 
According to this rule, only up to $k$ D3-branes can be suspended between the NS5-branes. 
This provides, in the pure Chern-Simons theory, 
the limitation $\frac Nk\le 1$ for the 't~Hooft coupling for which the supersymmetry is preserved. 
In the presence of $N_f$ D5-branes, D3-branes between the NS5-branes can be broken by these D5-branes. 
Therefore, more D3-branes can be present without breaking supersymmetry. 
The upper bound on $N$ is therefore 
\begin{equation}
N \le k+N_f, 
   \label{SUSY limit}
\end{equation}
implying 
\begin{equation}
\frac Nk \le 1+\frac{N_f}k. 
   \label{breaking}
\end{equation}
The inequality (\ref{SUSY limit}) can be also understood in the following way. 
One may start with a configuration in which all the $k+N_f$ D5-branes are intersecting to an NS5-brane. 
For this configuration, the condition for preserving the supersymmetry is (\ref{SUSY limit}). 
One then separates $N_f$ D5-branes from the intersection and obtains the configuration discussed above. 
This procedure consists of the web-deformation of the D5-NS5 system which does not break any supersymmetry. 
Therefore, the condition (\ref{SUSY limit}) is the condition for preserving the supersymmetry for the flavored Chern-Simons 
theory. 

\vspace{5mm}

It is interesting to see whether the Wilson loop in this flavored theory vanishes when the inequality (\ref{SUSY limit}) 
is saturated. 

The planar resolvent of ${\cal N}=3$ Chern-Simons theory with flavors was obtained in \cite{Suyama:2010hr}. 
The Wilson loop is obtained from the resolvent. 
The explicit expression is as follows, 
\begin{equation}
\langle W \rangle = \frac1{t_1}\left[ \frac{(a-a^{-1})^2}{4(2+a+a^{-1})}-t_2+\frac{t_2}2\sqrt{2+a+a^{-1}} \right], 
\end{equation}
where $t_1\propto \frac Nk$, $t_2\propto \frac{N_f}k$ and $a$ is determined by 
\begin{equation}
\log\frac4{2+a+a^{-1}}+t_2\sqrt{\frac4{2+a+a^{-1}}} = t_2-t_1. 
\end{equation}
It is convenient to introduce $u$ such that $a=e^{2u}$. 
Then we obtain 
\begin{eqnarray}
\langle W \rangle 
&=& \frac1{t_1}\Bigl[ \sinh^2u+t_2(\cosh u-1) \Bigr] \nonumber \\
&=& \frac1{t_1}\Bigl[ \sinh^2u+2t_2\sinh^2\frac u2 \Bigr], 
    \label{WL}
\end{eqnarray}
where $u$ satisfies 
\begin{equation}
-2\log\cosh u+\frac{t_2}{\cosh u} = t_2-t_1. 
\end{equation}
Note that $t_2=0$ implies $\cosh u=e^{t_1/2}$ which then implies 
\begin{equation}
\langle W \rangle = e^{t_1/2}\frac{\sinh(t_1/2)}{t_1/2}. 
\end{equation}
Recalling $t_1=2\pi i\lambda$, we found that the formula (\ref{WL}) reproduces the known result (\ref{CSWL}). 

The vanishing of the Wilson loop requires 
\begin{equation}
\cosh^2\frac u2 = -\frac{t_2}2, 
\end{equation}
or 
\begin{equation}
\cosh u = -1-t_2. 
\end{equation}
Therefore, the Wilson loop vanishes only when 
\begin{equation}
-\log(1+t_2)^2-\frac{t_2}{1+t_2} = t_2-t_1. 
\end{equation}
This has nothing to do with the supersymmetry breaking condition (\ref{breaking}). 

\vspace{5mm}

Note that the condition (\ref{breaking}) for the supersymmetry breaking is consistent with the result in \cite{Suyama:2010hr}. 
In \cite{Suyama:2010hr}, we started with GT theory whose gauge group is U$(N)_{k_1}\times$U$(N)_{-k_2}$. 
If $k_1=k_2$, then it is ABJM theory, and therefore the supersymmetry is preserved for any value of the parameters. 
Then, we took the limit $k_2\to\infty$. 
This limit is a weak coupling limit in terms of the second gauge group, and therefore, the supersymmetry should not be broken. 
The resulting theory is a Chern-Simons theory with gauge group U$(N)_{k_1}$ coupled to $N_f=2N$ fundamental 
hypermultiplets. 
This theory satisfies the condition (\ref{breaking}) for any choice of $N$ and $k_1$, so the supersymmetry is preserved. 
This must be the case since the localization formula used in \cite{Suyama:2010hr} 
is valid in the presence of the preserved supersymmetry.

\end{document}